\documentclass[pre,twocolumn]{revtex4}

\usepackage{amssymb,amsmath}

\usepackage{graphicx}
\usepackage{color}

\newcommand\beq{\begin{equation}}
\newcommand\eeq{\end{equation}}
\newcommand\beqa{\begin{eqnarray}}
\newcommand\eeqa{\end{eqnarray}}
\newcommand{\dd}{\text{d}}

\newcommand{\al}{\alpha}

\begin{document}

\title{Shear-rate dependent transport coefficients in granular suspensions}
\author{Vicente Garz\'o\email{vicenteg@unex.es} \homepage{http://www.eweb.unex.es/eweb/fisteor/vicente/}}
\affiliation{Departamento de F\'{\i}sica and Instituto de Computaci\'on Cient\'{\i}fica Avanzada (ICCAEx), Universidad de Extremadura, E-06071 Badajoz, Spain}

\begin{abstract}

A recent model for monodisperse granular suspensions is used to analyze transport properties in spatially inhomogeneous states close to the simple (or uniform) shear flow. The kinetic equation is based on the inelastic Boltzmann (for low density gases) with the presence of a viscous drag force that models the influence of the interstitial gas phase on the dynamics of grains. A normal solution is obtained via a Chapman-Enskog-like expansion around a (local) shear flow distribution which retains all the hydrodynamic orders in the shear rate. To first-order in the expansion, the transport coefficients characterizing momentum and heat transport around shear flow are given in terms of the solutions of a set of coupled linear integral equations which are approximately solved by using a kinetic model of the Boltzmann equation. To simplify the analysis, the steady-state conditions when viscous heating is compensated by the cooling terms arising from viscous friction and collisional dissipation are considered to get the explicit forms of the set of generalized  transport coefficients. The shear-rate dependence of some of the transport coefficients of the set is illustrated for several values of the coefficient of restitution.
\end{abstract}

\draft
\date{\today}
\maketitle

\section{Introduction}
\label{sec1}

Although in nature granular materials are usually immersed in a gas or liquid phase (like the air, for instance), the influence of the latter on the transport properties of solid particles is generally neglected in most theoretical and computational studies. However, high-velocity, gas-solid flows occur in a wide range of practical applications (like circulating fluidized beds, for instance) and hence, the impact of the gas phase on grains should be accounted for in many circumstances. An example corresponds to species segregation problems where several works \cite{MLNJ01,NSK03,SSK04,WZXS08,CPSK10,PGM14} have shown that the presence of the interstitial fluid may significantly change the segregation phase-diagrams obtained in previous studies for (dry) granular flows (namely, when the role of the gas phase is neglected).

At a kinetic theory level, the description of such multiphase flows is quite intricate since the system involves two different phases (solid particles and interstitial fluid) and hence, one would need to solve a set of two coupled kinetic equations for each one of the velocity distribution functions of the different phases. On the other hand, in order to gain some insight into this complex problem, most of the models proposed in the literature for gas-solid flows have considered a single kinetic equation for the solid particles where the effect of the surrounding fluid on them is taken into account through an effective external force $\mathbf{F}_\text{fluid}$ \cite{models}.

A simple and realistic way of modeling the fluid-solid interaction force $\mathbf{F}_\text{fluid}$ is by means of a viscous drag force given by
\beq
\label{1.1}
\mathbf{F}_{\text{fluid}}=-m \gamma (\mathbf{v}-\mathbf{U}_g),
\eeq
where $m$ and $\mathbf{v}$ are the mass and the velocity of the particles, respectively, $\gamma$ is the friction coefficient (assumed to be proportional to the gas viscosity $\mu_g$), and $\mathbf{U}_g$ is the (known) mean velocity of the gas phase. The model defined in Eq.\ \eqref{1.1} has been recently considered in different papers to study the shear rheology of frictional hard-sphere suspensions \cite{H13,SMMD13,WGZS14,MSMD15}. In addition, model \eqref{1.1} can be seen as a simplified version of the more general particle acceleration model proposed in Ref.\ \cite{GTSH12} where the effect of the gas phase is not only accounted for by the drag force  \eqref{1.1} but also by means of a Langevin-like term. This latter term takes into account the added effects coming from neighboring particles and can be neglected when the mean velocity of the solid particles follows the mean flow velocity of the gas ($\mathbf{U} \simeq \mathbf{U}_g$). Here, $\mathbf{U}$ [defined below  in Eq.\ \eqref{2.5}] denotes the mean flow velocity of the solid particles. Thus, the results derived from this simple version of the model can be considered of practical interest to analyze linear transport in dilute gas-solid flows when the mean flow velocity of the solid and gas phases are practically the same (like, for instance, in the simple or uniform shear flow (USF) state \cite{TK95,SMTK96,ChVG15}).

An interesting problem is to assess the impact of the interstitial fluid on the transport properties of solid particles under USF. As usual, solid particles are modeled as a gas of inelastic smooth hard spheres with a constant coefficient of restitution $0<\al\leq 1$. The USF state is likely the simplest flow problem since the only nonzero hydrodynamic gradient is $\partial U_x/\partial y \equiv a$, where $a$ is the \emph{constant} shear rate. Due to its simplicity, this state has been widely studied in the past for dry elastic \cite{GS03} and inelastic \cite{C90,G03} gases as an ideal testing ground to shed light on the response of the system to large shear rates. Years ago, two independent papers \cite{L06,G06} analyzed momentum and heat transport around USF for a dry \emph{dilute} granular gas in spatially inhomogeneous states close to the USF. The heat and momentum fluxes were determined to first order in the deviations of the hydrodynamic field gradients from their values in the reference USF state. Given that the granular gas is strongly sheared, the corresponding transport coefficients are nonlinear functions of both the shear rate and the coefficient of restitution $\al$. This is one of the main new added values of these constitutive equations. On the other hand, in order to get explicit results and due to the mathematical difficulties involved in the general non-stationary problem, a particular sort of perturbations were considered to obtain the generalized transport coefficients under steady state conditions. Given that the (scaled) shear rate $a^*\equiv a/\nu$ [$\nu$ is a collision frequency for hard spheres, see Eq.\ \eqref{2.24}] and $\al$ are coupled in the steady state, then the generalized transport coefficients are \emph{only} functions of the coefficient of restitution $\al$.

The aim of this paper is to study transport around USF in dilute granular suspensions. As said before, the starting point is the inelastic Boltzmann equation \cite{BP04,P15} with the presence of the viscous drag force \eqref{1.1}. As in Refs.\ \cite{L06,G06}, the Boltzmann equation is  solved by means of a Chapman-Enskog-like expansion \cite{CC70} around the reference USF distribution $f^{(0)}$. Since the latter applies for arbitrary values of the shear rate $a$, the successive approximations $f^{(k)}$ in the perturbation expansion retain all the hydrodynamic orders in $a$. Consequently, the problem deals with two kinds of spatial gradients: \emph{small} gradients due to perturbations of the USF and arbitrary shear rates due to the background shear flow. As in Refs.\ \cite{L06,G06}, the study here is restricted to first order in the spatial gradients in the density, temperature, and flow velocity. The question arises then as to whether, and if so to what extent, the conclusions drawn from Refs.\ \cite{L06,G06} may be altered when the new ingredient associated with the presence of the gas phase is accounted for in the theory.

In the first-order approximation, the momentum transport is characterized by the viscosity tensor $\eta_{ijk\ell}$ while the heat flux is characterized by the thermal conductivity tensor $\kappa_{ij}$ and the Dufour-like tensor $\mu_{ij}$. As in the case of dry granular gases, to get explicit analytical results, the steady state conditions are considered and hence, the (scaled) friction coefficient $\gamma^*$ (which characterizes the amplitude of the drag force) is given in terms of the (independent) relevant parameters $a^*$ and $\al$. This contrasts with the results offered in Refs.\ \cite{L06,G06} since the transport coefficients are now explicitly obtained as nonlinear functions of both the shear rate and the coefficient of restitution.

For ordinary fluids (elastic collisions), several previous works studied the shear-rate dependence of the thermal conductivity tensor under shear flow. Thus, Evans \cite{E91} derived years ago a Green-Kubo formula for the thermal conductivity in a strongly shearing fluid. In a similar way as in the equilibrium case, the thermal conductivity of a shearing steady state is expressed in terms of fluctuations in steady heat flux. This formula was subsequently employed to calculate the shear-rate dependence of the thermal conductivity of a Lennard-Jones fluid via nonequilibrium molecular dynamics simulations methods \cite{DE93}. In the context of kinetic theory, an explicit expression of the thermal conductivity tensor was derived \cite{G93,G95} by solving the Boltzmann equation by means of an expansion around the shear flow state. These analytical results were shown to compare qualitatively well with the computer simulations performed in Ref.\ \cite{DE93}. It must be noted that the calculations carried out in Refs.\ \cite{G93,G95} slightly differ from the ones carried out in this paper since the former require an additional external force to reach a steady state with constant pressure and linear shear field. Apart from these papers, a more recent paper \cite{SA14} for dry granular gases has determined the thermal conductivity tensor via an expansion around an anisotropic Gaussian distribution function. The authors derived a generalized Fourier law for the granular heat flux where the thermal conductivity is characterized by an anisotropic second rank tensor. A comparison between the results obtained here with those reported before for ordinary \cite{G93,G95} and granular \cite{SA14} sheared gases will be made in sec.\ \ref{sec7}.

The plan of the paper is as follows. In Sec.\ \ref{sec2} the Boltzmann kinetic equation is introduced and its corresponding balance equations derived. Section \ref{sec3} deals with the relevant results derived in the (unperturbed) USF problem by solving the Boltzmann equation by means of Grad's moment method \cite{G49}. In Sec.\ \ref{sec4} the problem we are interested in is described and the set of coupled linear equations defining the generalized coefficients $\eta_{ijk\ell}$, $\kappa_{ij}$, and $\mu_{ij}$ are provided. Explicit expressions for these shear-rate dependent transport coefficients are then obtained in Sec.\ \ref{sec5} by employing a kinetic model of the Boltzmann equation. The details of the calculations are displayed along several Appendices. The shear-rate dependence of some transport coefficients is illustrated in Sec.\ \ref{sec6} for different values of the coefficient of restitution. Finally, in Sec.\ \ref{sec7} the paper is closed with some concluding remarks.

\section{Boltzmann kinetic equation for monodisperse granular suspensions}
\label{sec2}


We consider a granular suspension of solid particles of mass $m$ and diameter $\sigma$ immersed in a gas of viscosity $\mu_g$. Under rapid flow conditions, particles are modeled as a gas of smooth hard spheres or disks with inelastic collisions. The inelasticity of collisions is characterized by a \emph{constant} (positive) coefficient of normal restitution $\al \leq 1$. As said in the Introduction, a simple and usual usual way of modeling the effect of the interstitial gas on the dynamic properties of the solid particles is through the presence of nonconservative external forces. These forces are incorporated into the corresponding Boltzmann kinetic equation of the solid particles. Thus, in the low-density regime, the one-particle velocity distribution function $f(\mathbf{r}, \mathbf{v},t)$ of grains obeys the kinetic equation \cite{BP04}
\beq
\label{2.1}
\frac{\partial f}{\partial t}+{\bf v}\cdot \nabla f+\frac{\partial}{\partial \mathbf{v}}\cdot \left(\frac{\mathbf{F}_{\text{fluid}}}{m}f\right)=J[\mathbf{v}|f,f],
\eeq
where the Boltzmann collision operator $J\left[{\bf v}|f,f\right]$ is given by
\beqa
\label{2.2}
J\left[{\bf v}_{1}|f,f\right]&=&\sigma^{d-1}\int \dd {\bf v}
_{2}\int \dd \widehat{\boldsymbol{\sigma}}\,\Theta (\widehat{{\boldsymbol {\sigma }}}
\cdot {\bf g}_{12})(\widehat{\boldsymbol {\sigma }}\cdot {\bf g}_{12})\nonumber\\
& & \times \left[\alpha^{-2}f({\bf v}_1')f({\bf v}_2')-f({\bf v}_1)f({\bf v}_2)\right].
\eeqa
Here, $d$ is the dimensionality of the system ($d=2$ for disks and $d=3$ for spheres), $\boldsymbol
{\sigma}=\sigma \widehat{\boldsymbol {\sigma}}$, $\widehat{\boldsymbol {\sigma}}$ being a unit vector pointing in the direction from the center of particle $1$ to the center of particle $2$, $\Theta $ is the Heaviside step function, and ${\bf g}_{12}={\bf v}_{1}-{\bf v}_{2}$ is the relative velocity. The primes on the velocities in Eq.\ \eqref{2.2} denote the initial
values $\{{\bf v}_{1}^{\prime}, {\bf v}_{2}^{\prime }\}$ that lead to
$\{{\bf v}_{1},{\bf v}_{2}\}$ following a binary collision:
\begin{equation}
{\bf v}_{1,2}^{\prime}={\bf v}_{1,2}\mp\frac{1}{2}\left( 1+\alpha^{-1}\right)
(\widehat{{\boldsymbol {\sigma }}}\cdot {\bf g}_{12})\widehat{{\boldsymbol {\sigma }}}.
\label{2.3}
\end{equation}

As mentioned in the Introduction, a simplest way of modeling the fluid-solid interaction force $\mathbf{F}_{\text{fluid}}$ is through the drag force \eqref{1.1} where
\begin{equation}
\label{2.5}
{\bf U}(\mathbf{r},t)=\frac{1}{n(\mathbf{r},t)}\int \;\dd{\bf v} \; {\bf v}\;  f(\mathbf{r},\mathbf{v},t)
\end{equation}
is the mean flow velocity of the solid particles, and
\beq
\label{2.6}
n(\mathbf{r},t)=\int\; \dd \mathbf{v}\; f(\mathbf{r},\mathbf{v},t)
\eeq
is the number density of particles. Thus, according to Eqs.\ \eqref{1.1} and \eqref{2.1}, the Boltzmann equation becomes
\beq
\label{2.4}
\frac{\partial f}{\partial t}+{\bf v}\cdot \nabla f-\gamma\Delta \mathbf{U}\cdot \frac{\partial f}{\partial \mathbf{V}}-\gamma\frac{\partial}{\partial \mathbf{V}}
\cdot \mathbf{V} f=J[\mathbf{v}|f,f],
\eeq
where $\Delta \mathbf{U}=\mathbf{U}-\mathbf{U}_g$, and ${\bf V}={\bf v}-{\bf U}$ is the peculiar velocity. Note that in the case of very dilute suspensions, $\gamma$ is assumed to be a constant \cite{K90,KS99,WKL03}.

The macroscopic balance equations for the densities of mass, momentum and energy can be obtained by multiplying Eq.\ \eqref{2.4} by $1$, $m \mathbf{V}$, and $\frac{1}{2}m V^2$, respectively, and integrating over velocity. The result is
\beq
\label{2.7}
D_t n+n \nabla \cdot \mathbf{U}=0,
\eeq
\beq
\label{2.8}
D_t \mathbf{U}+(m n)^{-1} \nabla \cdot \mathsf{P}=-\gamma\Delta \mathbf{U},
\eeq
\beq
\label{2.9}
D_tT+\frac{2}{d n}\left( \nabla \cdot \mathbf{q}+\mathbf{P}:\nabla \mathbf{U} \right)=-2T\gamma-T \zeta.
\eeq
Here, $D_t\equiv \partial_t+\mathbf{v}\cdot \nabla$ is the material derivative,
\beq
\label{2.9.1}
T(\mathbf{r},t)=\frac{m}{d n(\mathbf{r},t)}\int\; \dd \mathbf{v}\; V^2\; f(\mathbf{r},\mathbf{v},t),
\eeq
is the \emph{granular} temperature,
\beq
\label{2.10}
P_{ij}(\mathbf{r},t)=m\int\; \dd \mathbf{v} V_i V_j f(\mathbf{r},\mathbf{v}, t),
\eeq
is the pressure tensor,
\beq
\label{2.11}
\mathbf{q}(\mathbf{r},t)=\frac{m}{2}\int\; \dd \mathbf{v} V^2 \mathbf{V} f(\mathbf{r},\mathbf{v}, t),
\eeq
is the heat flux, and
\beq
\label{2.12}
\zeta(\mathbf{r},t)=-\frac{m}{d n(\mathbf{r},t) T(\mathbf{r},t)}\int\; \dd \mathbf{v} V^2 J[\mathbf{v}|f,f]
\eeq
is the cooling rate characterizing the rate of energy dissipated due to collisions.

Notice that the interaction of solid particles with the gas phase is modeled solely by the friction term \eqref{1.1} since the term accounting for the momentum transferred from the gas (bath) to the granular particles (which is modeled by a stochastic force) has been neglected for the sake of simplicity. This stochastic force contributes to the Boltzmann equation \eqref{2.4} with a Langevin-like term of the form $-\frac{1}{2}\xi \partial^2 f/\partial V^2$, where $\xi$ is the strength of the noise term. As said in sec.\ \ref{sec1}, this stochastic term was considered in the complete suspension model proposed in Ref.\ \cite{GTSH12}. For elastic collisions and zero shear rate, the inclusion of the above stochastic term yields the balance equation $\partial_t T=-2T \gamma+m\xi$ and so, the Boltzmann equation \eqref{2.4} admits a stable steady equilibrium state. Indeed, it is precisely the condition of admitting an equilibrium state  that gives rise to a  fluctuation-dissipation theorem \cite{M89} fixing the strength of the noise term [i.e., $\xi=2\gamma T/m$, where $T$ is the steady equilibrium temperature]. The omission of this Langevin-like term could be justified if the bath temperature is very low compared to the granular temperature or if the mean flow velocities of solid and gas phases are quite similar \cite{GTSH12}.

On the other hand, in spite of the absence of the Langevin-like term, the Boltzmann equation \eqref{2.4} still admits a simple solution in the homogenous state (zero shear rate) for elastic collisions ($\al=1$). This solution is given by a time-dependent Maxwellian distribution. For homogeneous states, Eq.\ \eqref{2.4} becomes
\beq
\label{2.12.1}
\frac{\partial f}{\partial t}-\gamma\frac{\partial}{\partial \mathbf{v}}
\cdot \mathbf{v} f=J[\mathbf{v}|f,f],
\eeq
where an appropriate selection of the frame of reference where  the mean flow velocity vanishes ($\mathbf{U}=\mathbf{U}_g=\mathbf{0}$) has been chosen.  The only relevant balance equation is that of the temperature \eqref{2.9} which reads
\beq
\label{2.12.2}
\frac{\partial \ln T}{\partial t}=-2 \gamma.
\eeq
Since $\gamma \equiv \text{const.}$, then the solution to Eq.\ \eqref{2.12.2} is simply
\beq
\label{2.12.3}
T(t)=T(0) e^{-2\gamma t},
\eeq
where $T(0)$ is the initial temperature. Under these conditions, it is easy to see that the Boltzmann equation \eqref{2.12.1} has the solution \cite{GSB90,PG14}
\beq
\label{2.12.4}
f_0(\mathbf{v},t)=n \left(\frac{m}{2\pi T(t)}\right)^{d/2} \exp \left(-\frac{m v^2}{2 T(t)}\right),
\eeq
where $T(t)$ is given by \eqref{2.12.3}. An H-theorem has been also proved \cite{GSB90} for the distribution $f_0$ in the sense that, starting from any initial condition and in the presence of the viscous drag force $\gamma \mathbf{v}$, the velocity distribution function $f(\mathbf{r}, \mathbf{v}, t)$ reaches in the long time limit the Maxwellian form \eqref{2.12.4} with a time-dependent temperature.

Before closing this Section, it is interesting to remark the situations in which the suspension model \eqref{2.4} is expected to provide reliable predictions. As has been previously discussed in several papers \cite{models,TK95,SMTK96,WKL03}, since the form of the Boltzmann collision operator \eqref{2.2} is the same as for a dry granular gas, one expects that the model \eqref{2.4} is appropriate for problems where the stresses applied by the gas phase on particles have only a weak influence on the dynamics of grains. This necessarily requires that the mean-free time between collisions is much shorter than the viscous relaxation time due to the viscous drag force. For other kind of systems (e.g., glass beads in liquid water), one should take into account the influence of the interstitial fluid on the Boltzmann collision operator.

\section{Simple shear flow problem in dilute granular suspensions}
\label{sec3}

We assume now that the suspension is in \emph{steady} USF. This state is macroscopically defined by a constant density $n$ and temperature $T$ and the mean velocity $\mathbf{U}$ is
\beq
\label{2.15}
U_i=a_{ij}r_j, \quad a_{ij}=a\delta_{ix}\delta_{jy},
\eeq
where $a$ is the constant shear rate. In addition, as usual in uniform sheared suspensions \cite{TK95,SMTK96,ChVG15}, the average velocity of particles follows the velocity of the gas phase and so, $\mathbf{U}=\mathbf{U}_g$. In this case, $\Delta {\bf U}=\textbf{0}$ and the Boltzmann equation \eqref{2.4} becomes
\beq
\label{2.16}
-aV_y\frac{\partial f}{\partial V_x}-\gamma\frac{\partial}{\partial
{\bf V}}\cdot {\bf V} f =J[\mathbf{V}|f,f].
\eeq
Upon writing Eq.\ \eqref{2.16} use has been made of the fact that the USF state becomes spatially uniform when one expresses the Boltzmann equation in terms of the peculiar velocity $V_i=v_i-a_{ij}r_j$ \cite{DSBR86}. In the USF problem, the heat flux vanishes and the only relevant balance equation is that of the temperature \eqref{2.9}. In the steady state and for the geometry of the USF, Eq.\ \eqref{2.9} reads
\beq
\label{2.17}
\frac{2}{d n} P_{xy} a=-2T\gamma-\zeta T.
\eeq
Equation \eqref{2.17} implies that the viscous heating term ($-aP_{xy}>0$) is exactly canceled by the cooling terms arising from viscous friction ($\gamma T$) and collisional dissipation ($\zeta T$). Thus, in stationary conditions, for a given value of $\gamma$, the (steady) temperature is a function of the shear rate $a$ and the coefficient of restitution $\al$. Equivalently, one might chose $\gamma$ and $\al$ as independent parameters instead of $a$ and $\al$. This was the choice made in Refs.\ \cite{TK95,SMTK96,ChVG15}. Since we are mainly interested here in obtaining the shear-rate dependence of the transport coefficients, the former choice will be considered in this paper. A remarkable point is that a steady state is still possible for suspensions when the collisions are elastic ($\al=1$ and so, $\zeta=0$) provided $\gamma=-P_{xy} a/(d p)$, where $p=n T$ is the hydrostatic pressure.

The USF state is non-Newtonian. This can be characterized by generalized transport coefficients measuring the departure of transport coefficients from their Navier-Stokes forms. Thus, one can define a non-Newtonian shear viscosity coefficient $\eta(\al, a)$ by
\beq
\label{2.18.0}
P_{xy}=-\eta(\al,a) a.
\eeq
Moreover, while $P_{xx}=P_{yy}=P_{zz}$ in the Navier-Stokes domain, normal stress differences are present in the USF state.

The elements of the pressure tensor $P_{ij}$ can be obtained by multiplying both sides of Eq.\ \eqref{2.16} by $mV_iV_j$ and integrating over velocity. The result is
\beq
\label{2.18}
a_{ik}P_{kj}+a_{jk}P_{ki}+2\gamma P_{ij}=\Lambda_{ij},
\eeq
where
\beq
\label{2.19}
\Lambda_{ij}\equiv \int\; \dd\mathbf{v}\; m V_iV_j J[\mathbf{V}|f,f].
\eeq
So far, the hierarchy \eqref{2.18} is still exact. However, the exact expression of the collision integral $\Lambda_{ij}$ is not known (even for elastic collisions). A good estimate of $\Lambda_{ij}$ can be obtained by using Grad's approximation to $f$ \cite{G49}, namely,
\beq
\label{2.20}
f(\mathbf{V})\to f_\text{M}(\mathbf{V}) \left(1 +\frac{m}{2nT^2}V_iV_j \Pi_{ij}\right),
\eeq
where
\begin{equation}
\label{2.21}
f_\text{M}(\mathbf{V})=n\left(\frac{m}{2\pi T}\right)^{d/2}e^{-mV^2/2T}
\end{equation}
is the local equilibrium distribution function, and
\begin{equation}
\label{2.22}
\Pi_{ij}=P_{ij}-p\delta_{ij}
\end{equation}
is the traceless part of the pressure tensor. When Eq.\ \eqref{2.20} is substituted into the definition of $\Lambda_{ij}$ and nonlinear terms in $\Pi_{ij}$ are neglected, one gets the result \cite{G02}
\beq
\label{2.23}
\Lambda_{ij}=-\nu \left(\beta \Pi_{ij}+\zeta^* P_{ij}\right),
\eeq
where
\beq
\label{2.24}
\nu=\frac{8}{d+2}\frac{\pi^{(d-1)/2}}{\Gamma\left(\frac{d}{2}\right)}n\sigma^{d-1}\sqrt{\frac{T}{m}}
\eeq
is an effective collision frequency,
\beq
\label{2.25}
\zeta^*=\frac{\zeta}{\nu}=\frac{d+2}{4d}\left(1-\alpha^2\right)
\end{equation}
is the dimensionless cooling rate evaluated in the local equilibrium approximation and
\beq
\label{2.26}
\beta=\frac{1+\al}{2}\left[1-\frac{d-1}{2d}(1-\al)\right].
\eeq
As will show below, the determination of the collisional moment $\Lambda_{ij}$ by considering only linear terms  yields $P_{xx}\neq P_{yy}$ but $P_{yy}=P_{zz}$. This latter identity disagrees with computer simulation results \cite{TK95,SMTK96,ChVG15}. The evaluation of $\Lambda_{ij}$ by retaining all the quadratic terms in the pressure tensor $P_{ij}$ has been recently carried out in Ref.\ \cite{ChVG15}. As expected, the addition of these nonlinear terms allows to evaluate the normal stress differences in the plane normal to the laminar flow (e.g., $P_{yy}-P_{zz}$). However, given that this difference is quite small, the expression \eqref{2.23} can be considered as a reliable approximation. Apart from its simplicity, the linear Grad solution is also essentially motivated by the desire of analytic expressions that show in a clean way the shear-rate dependence of the rheological properties.

Once the collisional moment $\Lambda_{ij}$ is known, the set of coupled equations for $P_{ij}$ can be easily solved. In terms of the reduced shear rate $a^*= a/\nu$ and the coefficient of restitution $\al$, the expressions for the (scaled) elements $P_{ij}^*=P_{ij}/p$ are
\beq
\label{2.27}
P_{yy}^*=P_{zz}^*=\frac{1}{1+2\chi}, \quad P_{xx}^*=d-(d-1)P_{yy}^*,
\eeq
\beq
\label{2.28}
P_{xy}^*=-\frac{\widetilde{a}}{(1+2\chi)^2},
\eeq
where $\widetilde{a}= a^*/\beta$, and $\chi$ is the real root of the cubic equation
\beq
\label{2.29}
\widetilde{a}^2=d \chi(1+2\chi)^2,
\eeq
namely,
\beq
\label{2.30}
\chi(\widetilde{a})=\frac{2}{3}\sinh^2\left[\frac{1}{6}\cosh^{-1}\left(1+\frac{27}{d}\widetilde{a}^2\right)\right].
\eeq
The (scaled) friction coefficient $\gamma^*=\gamma/\nu$ is defined as
\beq
\label{2.31}
\gamma^*=\beta\chi-\frac{1}{2}\zeta^*.
\eeq
In the case of elastic collisions ($\al=1$), Eqs.\ \eqref{2.27}--\eqref{2.31} agree with those obtained \cite{GS03} for a thermostatted dilute gas under USF. Moreover, the analytical results given by Eqs.\ \eqref{2.27}--\eqref{2.31} compare quite well with Monte Carlo simulations of the Boltzmann equation \cite{ChVG15}, even for strong inelasticity.

\begin{figure}
\includegraphics[width=0.7 \columnwidth,angle=0]{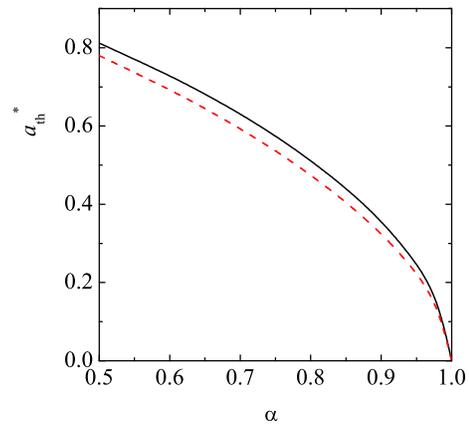}
\caption{(color online) Dependence of the threshold shear rate $a_{\text{th}}^*$ on the coefficient of restitution $\al$. The dashed line corresponds to a two-dimensional system ($d=2$) while the solid line refers to a three-dimensional ($d=3$) system. Points above the curves correspond to physical solutions $(\gamma^* \geq 0)$ while points below the curves refer to unphysical solutions $(\gamma^*<0$).
\label{fig1}}
\end{figure}

Since $\gamma^* \geq 0$, then necessarily $2\beta \chi-\zeta^* \geq 0$, according to Eq.\ \eqref{2.31}. This means that, at a given value of the coefficient of restitution, there is a threshold value of the (scaled) shear rate $a_{\text{th}}^*$ such that the steady state condition \eqref{2.17} admits a physical solution for $a^* \geq a_{\text{th}}^*$. This physical solution yields a positive granular temperature and is related to what Sangani \emph{et al.} \cite{SMTK96} call \emph{ignited} state. The value of $a_{\text{th}}^*$ is determined from the condition
\beq
\label{2.32}
2\beta \chi=\zeta^*.
\eeq
In particular, for elastic collisions, $\zeta^*=0$ and so, $a_{\text{th}}^*=0$. However, for inelastic collisions, $\zeta^*\neq 0$, and $a_{\text{th}}^*>0$. Thus, the rheological properties are only well-defined for shear rates beyond the nonvanishing $a_{\text{th}}^*$ in the case of granular suspensions ($\al \neq 1$). The $\al$-dependence of $a_{\text{th}}^*$ is plotted in Fig.\ \ref{fig1} for $d=2$ and $d=3$. For strong inelasticity, the curves highlight that the granular suspension is in general beyond the Navier-Stokes domain (non-Newtonian regime) since the (reduced) threshold shear rate $a_{\text{th}}^*$ is not small in general. Thus, for instance $a_{\text{th}}^*\simeq 0.512$ at $\al=0.8$ in the physical three-dimensional case.

\begin{figure}
\includegraphics[width=0.7 \columnwidth,angle=0]{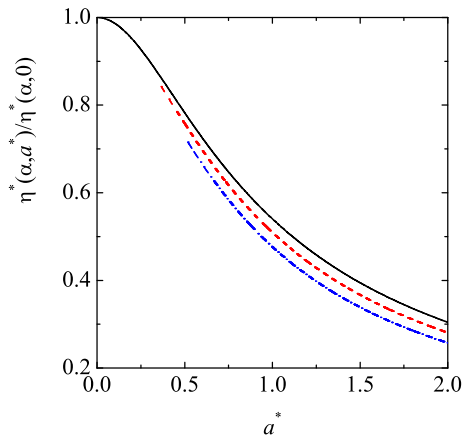}
\caption{(color online) Shear-rate dependence of the (scaled) generalized shear viscosity $\eta^*(\al,a^*)/\eta^*(\al, 0)$ for $d=3$ and three different values of the coefficient of restitution $\al$: $\al=1$ (solid line), $\al=0.9$ (dashed line), and $\al=0.8$ (dash-dotted line). Note that $a_{\text{th}}^*\simeq 0.359$ and $a_{\text{th}}^*\simeq 0.512$ for $\al=0.9$ and $\al=0.8$, respectively.
\label{fig2}}
\end{figure}

The fact that for granular suspensions ($\al \neq 1$) a steady state is only possible for sufficiently high shear rates can be easily understood from a physical point of view. For $\gamma^*=0$, the balance equation \eqref{2.32} establishes an intrinsic connection between the shear field [through the nonlinear function $\chi(\al,a^*)$] and the collisional dissipation [through the cooling rate $\zeta^*(\al)$] in the system. Thus, the magnitude of the (scaled) shear rate $a^*$ is set by the coefficient of restitution $\al$. Since $\zeta^* \propto 1-\al^2$, then the cooling rate increases with inelasticity. Moreover,  the rheological function $\chi$ increases with increasing $a^*$. Consequently, one needs to consider higher values of $a^*$ as $\al$ decreases to verify the condition \eqref{2.32} and achieve a steady state.

The (reduced) nonlinear shear viscosity $\eta^*=\eta/\eta_0$ can be easily identified from Eqs.\ \eqref{2.18.0} and \eqref{2.28}. Here, $\eta_0=p/\nu$ is the Navier-Stokes shear viscosity of an ordinary (elastic) gas of hard spheres. The expression of $\eta^*$ is given by
\beq
\label{2.33}
\eta^*(\al,a^*)=\frac{1}{\beta(1+2\chi)^2}.
\eeq
Since $\chi \sim a^{*2/3}$ for very large shear rates, then $\eta^* \sim a^{*-4/3}$ and goes to zero in the limit $a^*\to \infty$. To illustrate the shear-rate dependence of $\eta^*$, Fig.\ \ref{fig2} shows the ratio $\eta^*(\al,a^*)/\eta^*(\al,0)$ versus $a^*$ for $d=3$ and three different values of the coefficient of restitution $\al$. As mentioned before, except for elastic collisions and although $\eta^*$ is well defined for shear rates smaller than the threshold value $a_{\text{th}}^*$, the curves in Fig. \ \ref{fig2} start from the point $a^*=a_{\text{th}}^*$ for $\al \neq 1$. It appears that shear thinning (viscosity decreases with increasing shear rate) is always present, regardless of the value of the coefficient of restitution. We also observe that, at a given value of $a^*$, inelasticity inhibits the momentum transport. However, the influence of inelasticity on the (scaled) shear viscosity is not quantitatively significant.

\section{Transport coefficients for states close to USF}
\label{sec4}

Let us assume that we perturb the USF by small spatial gradients. This will give rise to new contributions to the momentum and heat fluxes that can be characterized by generalized transport coefficients. Since the system is strongly sheared, the corresponding transport coefficients are highly nonlinear functions of the shear rate. The evaluation of these coefficients is the main objective of the present paper.

As in previous papers \cite{L06,G06,G07}, in order to analyze this problem one has to start from the Boltzmann
equation \eqref{2.4} with a general time and space dependence.
First, it is convenient to continue using the relative velocity ${\bf V}={\bf v}-{\bf U}_0$, where
${\bf U}_0={\sf a}\cdot {\bf r}$ is
the flow velocity of the {\em unperturbed} USF state. As said before, the
only nonzero element of the tensor $\mathsf{a}$ is
$a_{ij}=a\delta_{ix}\delta_{jy}$. On the other hand, in the {\em perturbed} state the true velocity ${\bf U}$ is in general
different from ${\bf U}_0$ since ${\bf U}={\bf U}_0+\delta {\bf U}$, $\delta {\bf U}$ being a small perturbation to
${\bf U}_0$. As a consequence, the true peculiar velocity is now  ${\bf c}\equiv {\bf v}-{\bf U}={\bf V}-\delta{\bf U}$. In addition, for the sake of simplicity, we also assume that the interstitial gas is not perturbed and hence, $\mathbf{U}_g=\mathbf{U}_0$. Thus, in the Lagrangian frame moving with velocity $\mathbf{U}_0$, the convective operator $\mathbf{v}\cdot \nabla$ can be written as
\beq
\label{3.0}
\mathbf{v}\cdot \nabla f=\left(\mathbf{V}+\mathbf{U}_0\right)\cdot \nabla f=
-aV_y\frac{\partial f}{\partial V_x}+\left(\mathbf{V}+\mathbf{U}_0\right)\cdot \mathbf{\nabla}f,
\eeq
where the derivative $\nabla f$ is taken now at constant $\mathbf{V}$. In this case,
the Boltzmann equation \eqref{2.4} reads
\beq
\label{3.1}
\partial_{t}f-aV_y\frac{\partial f}{\partial V_x}+\left(\mathbf{V}+\mathbf{U}_0\right)\cdot \mathbf{\nabla}f
-\gamma \frac{\partial}{\partial{\bf V}}\cdot {\bf V} f
=J\left[{\bf v}|f,f\right].
\end{equation}
The corresponding macroscopic balance
equations associated with this disturbed USF state follows from the general equations (\ref{2.9})--(\ref{2.11})
when one takes into account that ${\bf U}={\bf U}_0+\delta {\bf U}$. The result is
\begin{equation}
\label{3.2}
\partial_tn+{\bf U}_0\cdot \nabla n=-\nabla \cdot (n\delta {\bf U}),
\end{equation}
\begin{equation}
\label{3.3}
\partial_t\delta {\bf U}+{\sf a}\cdot \delta {\bf U}+({\bf U}_0+\delta {\bf U})\cdot \nabla \delta {\bf U}
=-\gamma\delta \mathbf{U}-(mn)^{-1}\nabla \cdot {\sf P},
\end{equation}
\beqa
\label{3.4}
& & \frac{d}{2}n\partial_tT+\frac{d}{2}n({\bf U}_0+\delta {\bf U})\cdot \nabla T+aP_{xy}+\nabla
\cdot {\bf q}\nonumber\\
& & +{\sf P}:\nabla \delta {\bf U} =-\frac{d}{2}p\left(2\gamma+\zeta\right),
\eeqa
where the pressure tensor ${\sf P}$, the heat flux ${\bf q}$ and
the cooling rate $\zeta$ are defined by Eqs.\ (\ref{2.10})--(\ref{2.12}), respectively, with the replacement
${\bf V}\rightarrow {\bf c}$.

Since we are interested here in states close to the USF state, it is assumed that the deviations from the USF state are small and hence, the spatial gradients of $n$, $\delta \mathbf{U}$, and $T$ are small. In this case, Eq.\ \eqref{3.1} can be solved by means of a generalization of the conventional Chapman-Enskog method \cite{CC70}, where the velocity distribution function is expanded around a \emph{local} shear flow reference state in terms the small spatial gradients of the hydrodynamic fields relative to those of USF. This type of Chapman-Enskog-like expansion has been carried out for elastic gases to obtain the set of shear-rate dependent transport coefficients \cite{GS03,LD97} in a thermostatted shear flow problem and it has also been employed in the context of dry granular gases \cite{L06,G06,G07}.

The Chapman-Enskog method assumes the existence of a \emph{normal} solution in which all space and time dependence of the distribution function occurs through a functional dependence of the hydrodynamic fields
\beq
\label{3.4.1}
A(\mathbf{r},t)\equiv \left\{n(\mathbf{r},t), \delta \mathbf{U}(\mathbf{r},t), T(\mathbf{r},t)\right\}.
\eeq
This solution expresses the fact that the space dependence of the shear flow is absorbed in $\mathbf{V}$ and the remaining space and time dependence is through a functional dependence on the fields $A(\mathbf{r},t)$. As in the conventional Chapman-Enskog method, this functional dependence can be made local by an expansion of $f$ in powers of spatial gradients:
\begin{equation}
\label{3.5}
f({\bf r}, {\bf V},t)=f^{(0)}(A({\bf r}, t), {\bf V})+ f^{(1)}(A({\bf r}, t), {\bf V})+\cdots,
\end{equation}
where the reference zeroth-order distribution function corresponds
to the USF distribution function but taking into account the local
dependence of the density and temperature and the change ${\bf V}\rightarrow {\bf V}-\delta{\bf U}({\bf r}, t)$. The successive approximations $f^{(k)}$ are of order $k$ in the gradients of $n$, $T$, and
$\delta {\bf U}$ but retain all the orders in the shear rate $a$.
This is the main feature of this expansion. In addition, as in previous works \cite{GFHY16}, since the friction coefficient $\gamma$ does not induce any flux in the system, it is assumed then to be at least of zeroth order in the gradients. In this paper, only the first order approximation will be considered.

The expansion (\ref{3.5}) yields the corresponding expansion for the fluxes and the cooling rate when
one substitutes (\ref{3.5}) into their definitions (\ref{2.10})--(\ref{2.12}):
\begin{equation}
\label{3.6}
{\sf P}={\sf P}^{(0)}+{\sf P}^{(1)}+\cdots, \quad {\bf
q}={\bf q}^{(0)}+{\bf q}^{(1)}+\cdots,
\eeq
\beq
\label{3.6.1}
\zeta=\zeta^{(0)}+\zeta^{(1)}+\cdots.
\end{equation}
Finally, as in the usual Chapman-Enskog method, the time derivative is also expanded as
\begin{equation}
\label{3.7}
\partial_t=\partial_t^{(0)}+\partial_t^{(1)}+\partial_t^{(2)}+\cdots,
\end{equation}
where the action of each operator $\partial_t^{(k)}$ is obtained from the hydrodynamic equations
(\ref{3.2})--(\ref{3.4}). These results provide the basis for generating the Chapman-Enskog
solution to the inelastic Boltzmann equation (\ref{3.1}).

\subsection{Zeroth-order approximation}

Substituting the expansions (\ref{3.5})--(\ref{3.7}) into Eq.\ (\ref{3.1}), the kinetic equation
for $f^{(0)}$ is given by
\begin{equation}
\label{3.8}
\partial_t^{(0)}f^{(0)}-aV_y\frac{\partial}{\partial V_x}f^{(0)}
-\gamma \frac{\partial}{\partial{\bf V}}\cdot {\bf V} f^{(0)}
=J[{\bf V}|f^{(0)},f^{(0}].
\end{equation}
To lowest order in the expansion the conservation laws are
\begin{equation}
\label{3.10}
\partial_t^{(0)}n=0,\quad \partial_t^{(0)}T=-\left(\frac{2}{dn}a P_{xy}^{(0)}+2T\gamma+T\zeta^{(0)}\right),
\end{equation}
\begin{equation}
\label{3.11}
\partial_t^{(0)}\delta U_i=-a_{ij} \delta U_j-\gamma\delta U_i.
\end{equation}

As discussed in previous works \cite{L06,G06,G07}, for given values of $a$, $\gamma$ and $\alpha$, the steady
state condition (\ref{2.17}) establishes a mapping between the
density and temperature so that every density corresponds to one
and only one temperature. Since the density $n({\bf r}, t)$ and
temperature $T({\bf r}, t)$ are specified separately in the {\em
local} USF state, the viscous heating only partially compensates
for the collisional cooling and friction viscous dissipation and so, $\partial_t^{(0)} T \neq 0$.
Consequently, the zeroth-order distribution $f^{(0)}$ depends on
time through its dependence on the temperature and the (dimensionless) parameters $a^*$, $\gamma^*$, and $\al$ must be considered as independent parameters for general infinitesimal perturbations around the USF state. The fact that the temperature must be considered as a time-dependent parameter has been already accounted for in previous perturbation solutions around driven non-steady states \cite{GMT13,GChV13}.

Since $f^{(0)}$ is a normal solution, then
\begin{eqnarray}
\label{3.12}
\partial_t^{(0)}f^{(0)}&=&\frac{\partial f^{(0)}}{\partial
n}\partial_t^{(0)} n+\frac{\partial f^{(0)}}{\partial
T}\partial_t^{(0)} T+\frac{\partial f^{(0)}}{\partial \delta
U_i}\partial_t^{(0)} \delta U_i\nonumber\\
&=&-\left(\frac{2}{d n}a
P_{xy}^{(0)}+2T\gamma+T\zeta^{(0)}\right)\frac{\partial f^{(0)}}{\partial T}
\nonumber\\
& &-\left(a_{ij}\delta U_j
+\gamma\delta U_i \right)\frac{\partial f^{(0)}}{\partial \delta U_i}\nonumber\\
&=&-\left(\frac{2}{d n}a
P_{xy}^{(0)}+2T\gamma+T\zeta^{(0)}\right)\frac{\partial f^{(0)}}{\partial T}\nonumber\\
& & +\left(a_{ij}\delta U_j
+\gamma\delta U_i \right)\frac{\partial f^{(0)}}{\partial c_i}.
\end{eqnarray}
Upon deriving the last step in Eq.\ \eqref{3.12} use has been made of the fact that $f^{(0)}$ depends on $\delta {\bf U}$ only through the peculiar velocity ${\bf c}$. Substituting Eq.\ \eqref{3.12} into Eq.\ \eqref{3.8} yields the following kinetic equation for $f^{(0)}$:
\beqa
\label{3.13}
& & -\left(\frac{2}{d n}a
P_{xy}^{(0)}+2T\gamma+T\zeta^{(0)}\right)\frac{\partial f^{(0)}}{\partial T}
-ac_y\frac{\partial f^{(0)}}{\partial c_x}\nonumber\\
& &-\gamma \frac{\partial}{\partial{\bf c}}\cdot {\bf c} f^{(0)}
=J[{\bf V}|f^{(0)},f^{(0}].
\eeqa

The zeroth-order solution leads to ${\bf q}^{(0)}={\bf 0}$ by symmetry. The closed set of equations defining the zeroth-order pressure tensor $\mathsf{P}^{(0)}$ can be obtained from Eq.\ \eqref{3.13} by taking into account Eq.\ \eqref{2.23}. The result is
\beqa
\label{3.14}
& &
-\left(\frac{2}{d n}a
P_{xy}^{(0)}+2T\gamma+T\zeta^{(0)}\right)\frac{\partial P_{ij}^{(0)}}{\partial T} +
a_{ik}P_{jk}^{(0)}\nonumber\\
& & +a_{jk}P_{ik}^{(0)}+2\gamma P_{ij}^{(0)}=
-\nu\left[\beta \left(P_{ij}^{(0)}-p\delta_{ij}\right)+\zeta_0^*P_{ij}^{(0)}\right],
\nonumber\\
\eeqa
where $\zeta_0^*\equiv \zeta^{(0)}/\nu$ is defined by Eq.\ \eqref{2.25}.

The steady state solution of Eq.\ (\ref{3.14}) is given by Eqs.\ (\ref{2.27})--
(\ref{2.29}). However, for non-steady conditions, in general Eqs.\ \eqref{3.14} must be solved numerically to get
the dependence of the zeroth-order pressure tensor $P_{ij}^{(0)}(T)$ on temperature. In the hydrodynamic regime, it is expected that $P_{ij}^{(0)}$ adopts the form
\beq
\label{3.15}
P_{ij}^{(0)}=p P_{ij}^*(\gamma^*,a^*),
\eeq
where the temperature dependence of the (dimensionless) pressure tensor $P_{ij}^*$ is through its dependence on $\gamma^*$ and $a^*$. Since $\gamma^*\propto T^{-1/2}$ and $a^*\propto T^{-1/2}$, then
\beq
\label{3.16}
T\partial_T P_{ij}^{(0)}=P_{ij}^{(0)}-\frac{1}{2}p\left(\gamma^*\frac{\partial P_{ij}^*}{\partial \gamma^*}
+a^*\frac{\partial P_{ij}^*}{\partial a^*}\right).
\eeq
As it will be shown below, to determine the generalized transport coefficients in the steady state, one needs to know the derivatives $\partial_{\gamma^*}P_{ij}^*$ and $\partial_{a^*}P_{ij}^*$ in this state. These derivatives are evaluated in  Appendix \ref{appA}. In what follows, $P_{ij}^{(0)}(T)$ will be considered as a known function of $T$.
\begin{figure}
\includegraphics[width=0.7 \columnwidth,angle=0]{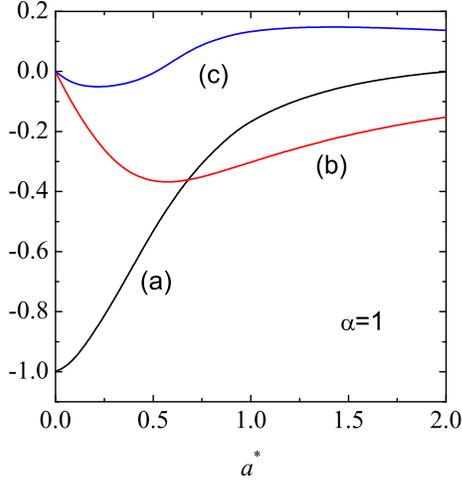}
\caption{(color online) Shear-rate dependence of the derivatives of the pressure tensor with respect to $a^*$ and $\gamma^*$ in the steady state for $d=3$ and $\al=1$. The lines (a), (b), and (c) correspond to $\partial_{a^*} P_{xy}^*=\partial_{\gamma^*} P_{xy}^*$, $\partial_{a^*} P_{yy}^*$, and $\partial_{\gamma^*} P_{yy}^*$, respectively.
\label{fig2.1}}
\end{figure}

The shear-rate dependence of the derivatives of the (reduced) pressure tensor with respect to $a^*$ and $\gamma^*$ in the steady state are illustrated in Fig.\ \ref{fig2.1} for a three-dimensional suspension with elastic collisions ($\al=1$). Although we could not analytically prove  the identity $\partial_{a^*} P_{xy}^*=\partial_{\gamma^*} P_{xy}^*$, numerical results systematically show this result. Since the magnitude of these derivatives is not in general quite small, it appears that their influence on transport cannot be in principle neglected.

\subsection{First-order approximation}

The first order approximation is worked out in Appendix \ref{appB}. Only the final results are given here. The velocity distribution function $f^{(1)}$ is
\begin{equation}
\label{3.17}
f^{(1)}={\bf X}_{n}\cdot \nabla n+ {\bf X}_{T}\cdot \nabla T+{\sf X}_{u}:\nabla \delta {\bf u},
\end{equation}
where the vectors ${\bf X}_{n}$ and ${\bf X}_{T}$ and the tensor ${\sf X}_{u}$ are the solutions of the following set of coupled linear integral equations:
\beqa
\label{3.18}
& & -\left(\frac{2}{d p}a
P_{xy}^{(0)}+2\gamma+\zeta^{(0)}\right)T \partial_T X_{n,i}- a
c_y\frac{\partial X_{n,i}}{\partial c_x}
\nonumber\\
& &
-\gamma \frac{\partial}{\partial{\bf c}}\cdot \left({\bf c}
X_{n,i}\right)+{\cal L}X_{n,i}+\frac{T}{n}\left[\frac{2a}{d p}(1-n\partial_n)
P_{xy}^{(0)}\right.\nonumber\\
& &\left. -\zeta^{(0)}\right]X_{T,i}
=Y_{n,i},
\eeqa
\beqa
\label{3.19}
& & -\left(\frac{2}{d p}a
P_{xy}^{(0)}+2\gamma+\zeta^{(0)}\right)T \partial_T X_{T,i}+
\left[\frac{2a}{d n}(\partial_T P_{xy}^{(0)})\right.\nonumber\\
& & \left.+2\gamma+\frac{3}{2}\zeta^{(0)}\right]X_{T,i}- a
c_y\frac{\partial X_{T,i}}{\partial c_x}-\gamma \frac{\partial}{\partial{\bf c}}\cdot \left({\bf c}
X_{T,i}\right)
\nonumber\\
& &
+{\cal L}X_{T,i}=Y_{T,i},
\eeqa
\beqa
\label{3.20}
& & -\left(\frac{2}{d p}a
P_{xy}^{(0)}+2\gamma+\zeta^{(0)}\right)T \partial_T X_{u,k\ell}
- a c_y\frac{\partial X_{u,k\ell}}{\partial c_x}\nonumber\\
& & -\gamma \frac{\partial}{\partial{\bf c}}\cdot \left({\bf c}
X_{u,k\ell}\right)+{\cal L}X_{u,k\ell}
-a\delta_{ky}X_{u,x\ell}\nonumber\\
& &-\gamma X_{u,k\ell}-\zeta_{u,k\ell}T\partial_T
f^{(0)}=Y_{u,k\ell},
\eeqa
where ${\bf Y}_n({\bf c})$, ${\bf Y}_T({\bf c})$, and ${\sf Y}_u({\bf c})$ are defined by Eqs.\ (\ref{b9})--(\ref{b11}),
respectively, and $\zeta_{u,k\ell}$ is defined by Eq.\ (\ref{b12}). An approximate expression of $\zeta_{u,k\ell}$ is given by Eq.\ (\ref{b13}). In addition, ${\cal L}$ is the linearized Boltzmann collision operator around the USF state, namely,
\begin{equation}
\label{3.21}
{\cal L}X \equiv -\left(J[f^{(0)},X]+J[X,f^{(0)}]\right).
\end{equation}
Note that, due to the presence of $P_{xy}^{(1)}$ in Eq.\ \eqref{b5}, the unknown coefficients $\eta_{xyk\ell}$ appear in the quantity $Y_{u,k\ell}$ of Eq.\ \eqref{3.20}. In the particular case of $\gamma^*=0$, Eqs.\ \eqref{3.18}--\eqref{3.20} are consistent with the results derived in Ref.\ \cite{G06} for dry granular gases.

With the distribution $f^{(1)}$ determined by Eq.\ \eqref{3.17}, the first-order corrections to the fluxes are given by
\begin{equation}
\label{3.22}
P_{ij}^{(1)}=-\eta_{ijk\ell} \frac{\partial \delta U_k}
{\partial r_{\ell}},
\end{equation}
\begin{equation}
\label{3.23}
q_i^{(1)}=-\kappa_{ij}\frac{\partial T}{\partial r_j}-
\mu_{ij}\frac{\partial n}{\partial r_j},
\end{equation}
where
\begin{equation}
\label{3.24}
\eta_{ijk\ell}=-\int\; \dd {\bf c}\, mc_ic_j X_{u,k\ell}({\bf c}),
\end{equation}
\begin{equation}
\label{3.25}
\kappa_{ij}=-\int\; \dd {\bf c}\, \frac{m}{2}c^2c_i X_{T,j}({\bf c}),
\end{equation}
\begin{equation}
\label{3.26}
\mu_{ij}=-\int\; \dd {\bf c}\, \frac{m}{2}c^2c_i X_{n,j}({\bf c}).
\end{equation}
Upon writing Eqs.\ \eqref{3.22}--\eqref{3.26} use has been made of
the symmetry properties of $X_{n,i}$, $X_{T,i}$, and $X_{u,ij}$.

In the absence of gas phase ($\gamma^*=0$), for $a^*=0$ and $\al=1$, the conventional Navier-Stokes constitutive equations for ordinary gases are reobtained, namely,
\beq
\label{3.26.1}
\eta_{ijk\ell}\to \eta_0 \left(\delta_{ik}\delta_{j\ell}+\delta_{jk}\delta_{i\ell}-\frac{2}{d}\delta_{ij}\delta_{k\ell}\right),
\eeq
\beq
\label{3.26.2}
\kappa_{ij}\to \kappa_0 \delta_{ij}, \quad \mu_{ij} \to 0.
\eeq
Here, $\eta_0=p/\nu$ and $\kappa_0=d(d+2)\eta_0/2(d-1)m$ are the expressions of the shear viscosity and thermal conductivity coefficients, respectively, of an ordinary gas of disks ($d=2$) or hard ($d=3$) spheres \cite{CC70}. In the absence of shear rate, the expressions of the Navier-Stokes coefficients of a granular suspension have been recently derived in Ref.\ \cite{GFHY16}.

In general, the set of {\em generalized} transport coefficients
$\eta_{ijk\ell}$, $\kappa_{ij}$, and $\mu_{ij}$ are nonlinear
functions of the coefficient of restitution $\alpha$, the
reduced shear rate $a^*$ and the reduced friction coefficient $\gamma^*$. The anisotropy induced in the system by
the shear flow gives rise to new transport coefficients,
reflecting broken symmetry. Since $P_{ij}^{(1)}$ is a symmetric and traceless tensor, then the viscosity tensor $\eta_{ijk\ell}$ is symmetric and traceless in $ij$, namely,
\beq
\label{3.26.1}
\eta_{ijk\ell}=\eta_{jik\ell}\neq \eta_{ij\ell k}, \quad \eta_{xxk\ell}+\eta_{yyk\ell}+\eta_{zzk\ell}+\cdots=0.
\eeq
The heat flux is expressed in terms of a thermal conductivity tensor $\kappa_{ij}$ and a Dufour-like
tensor $\mu_{ij}$. While the diagonal elements of both tensors can be interpreted as generalizations of the Navier-Stokes transport coefficients, the off-diagonal elements $\kappa_{xy}$, $\kappa_{yx}$, $\mu_{xy}$, and $\mu_{yx}$ are generalizations of Burnett coefficients that, for small shear rates, are proportional to $a^*$. In addition, because of symmetry reasons, the off-diagonal elements $xz$, $zx$, $yz$, and $zy$ of the tensors $\kappa_{ij}$ and $\mu_{ij}$ are identically zero. This is consistent with Eqs.\ \eqref{3.18} and \eqref{3.19}. The above behavior implies that if the thermal gradient is parallel to the $z$ axis ($\nabla T \parallel \widehat{\boldsymbol{z}}$), then $\mathbf{q}^{(1)} \parallel \widehat{\boldsymbol{z}}$, while if $\nabla T \perp \widehat{\boldsymbol{z}}$, then $\mathbf{q}^{(1)} \perp \widehat{\boldsymbol{z}}$. Similarly, many of the elements of the viscosity tensor $\eta_{ijk\ell}$ are zero. For instance, if the only nonzero velocity gradient is $\partial \delta U_x/\partial z$, then $P_{ij}^{(1)}=P_{xz}^{(1)}(\delta_{ix}\delta_{jz}+\delta_{jx}\delta_{iz})$.

\subsection{Steady state conditions}

As in the case of dry granular gases ($\gamma^*=0$), the evaluation of the transport coefficients $\eta_{ijk\ell}$, $\kappa_{ij}$ and $\mu_{ij}$ for general unsteady conditions is quite intricate. This is due essentially to the fact that the temperature dependence of the velocity moments of the distribution $f^{(0)}$ must be numerically determined. Thus, since we want to get analytical expressions for those coefficients, the present study is limited to steady state conditions. This means that the relation \eqref{2.17} is considered at the end of the calculations. In this state, the (scaled) shear rate $a^*$ is coupled to the (reduced) friction coefficient $\gamma^*$ and the coefficient of restitution $\alpha$ so that, only two of the three parameters are independent. Here, as alluded to in Sec.\ \ref{sec2}, $a^*$ and $\alpha$ are chosen as the independent (input) parameters of the problem. This allows us to independently assess the influence of shearing and inelasticity on momentum and heat transport. This contrasts with the analysis of dry granular gases \cite{G06} where both $a^*$ and $\al$  are considered as dependent parameters in the steady state.

Since the relation \eqref{2.17} holds in the steady state, the first term on the left hand side of the integral equations \ \eqref{3.18}--\eqref{3.20} vanishes. In this case, these equations become
\beqa
\label{3.28}
& & - a
c_y\frac{\partial X_{n,i}}{\partial c_x}-\gamma \frac{\partial}{\partial{\bf c}}\cdot \left({\bf c}
X_{n,i}\right)+{\cal L}X_{n,i}
\nonumber\\
& &
+\frac{T}{n}\left[\frac{2a}{d p}(1-n\partial_n)
P_{xy}^{(0)}-\zeta^{(0)}\right]X_{T,i}
=Y_{n,i},
\eeqa
\beqa
\label{3.29}
& &
\left[\frac{2a}{d n}(\partial_T P_{xy}^{(0)})+2\gamma+\frac{3}{2}\zeta^{(0)}\right]X_{T,i}- a
c_y\frac{\partial X_{T,i}}{\partial c_x}\nonumber\\
& & -\gamma \frac{\partial}{\partial{\bf c}}\cdot \left({\bf c}
X_{T,i}\right)
+{\cal L}X_{T,i}=Y_{T,i},
\eeqa
\beqa
\label{3.30}
& &
- a c_y\frac{\partial X_{u,k\ell}}{\partial c_x}-\gamma \frac{\partial}{\partial{\bf c}}\cdot \left({\bf c}
X_{u,k\ell}\right)+{\cal L}X_{u,k\ell}
-a\delta_{ky}X_{u,x\ell}\nonumber\\
& &-\gamma X_{u,k\ell}-\zeta_{u,k\ell}T\partial_T
f^{(0)}=Y_{u,k\ell}.
\eeqa
In Eqs.\ \eqref{3.28}--\eqref{3.30} it is understood that all the quantities are evaluated in the steady state. Moreover, the dependence of $P_{ij}^{(0)}$ on the temperature $T$ is given by Eq.\ \eqref{3.16} while the dependence of $P_{ij}^{(0)}$ on the density can be written as
\beq
\label{3.27}
n\partial_n P_{ij}^{(0)}=P_{ij}^{(0)}-p\left(\gamma^*\frac{\partial P_{ij}^*}{\partial \gamma^*}
+a^*\frac{\partial P_{ij}^*}{\partial a^*}\right).
\eeq

\section{Results from a BGK-like kinetic model}
\label{sec5}

Needless to say, the explicit form of the generalized transport coefficients $\eta_{ijk\ell}$, $\kappa_{ij}$, and $\mu_{ij}$ requires to solve the integral equations \eqref{3.28}--\eqref{3.30}. Apart from the mathematical
difficulties embodied in the Boltzmann collision operator ${\cal L}$, it is quite apparent that the fourth-degree velocity moments of the zeroth-order distribution $f^{(0)}$ are also needed to determine the heat flux transport coefficients $\mu_{ij}$ and $\kappa_{ij}$. Although these moments could in principle be determined from Grad's moment method by including them in the trial distribution \eqref{2.20}, their evaluation would be an intricate task.

A possible alternative could be the use of the so-called inelastic Maxwell models \cite{BCG00,BK00,EB02}, i.e., models for which the collision rate is independent of the relative velocity of the two colliding particles. The use of these models allows to obtain the velocity moments of the Boltzmann collision operator without the explicit knowledge of the velocity distribution function. This was the route followed in Ref.\ \cite{G07} to determine the shear-rate dependent transport coefficients in a dry granular sheared gas. However, apart from the difficulties associated with the evaluation of the fourth-degree moments and their derivatives, the results obtained for inelastic Maxwell models \cite{G07} show significant discrepancies from those obtained for inelastic hard spheres \cite{G06}.

Therefore, as in the previous study carried out for dry granular gases \cite{G06}, a model kinetic equation of the Boltzmann equation is considered to achieve explicit results. As for elastic collisions, the idea is to replace the true Boltzmann collision operator with a simpler, more tractable operator that retains the most relevant physical properties of the Boltzmann operator. Here, we consider a kinetic model \cite{BDS99} based on the well-known Bhatnagar-Gross-Krook (BGK) \cite{GS03} for
ordinary gases where the operator $J[f,f]$ is \cite{note}
\begin{equation}
\label{4.1}
J[f,f]\to -\beta \nu (f-f_\text{M})+\frac{\zeta}{2}\frac{\partial}
{\partial {\bf c}}\cdot \left({\bf c}f\right).
\end{equation}
Here, $\nu$ is the effective collision frequency defined by Eq.\ \eqref{2.24}, $f_\text{M}(\mathbf{c})$ is the Maxwellian distribution \eqref{2.21}, $\beta$ is given by Eq.\ \eqref{2.26} and $\zeta$ is the cooling rate. It is easy to see that the BGK model yields the same expressions for the pressure tensor in the steady USF state than those derived from Grad's method [Eqs.\ \eqref{2.27}--\eqref{2.29}]. Moreover, the fourth-degree velocity moments obtained from the BGK model compare quite well with Monte Carlo simulations \cite{ChVG15,AS05} of the Boltzmann equation. This confirms again the reliability of kinetic models to evaluate the velocity moments of the true Boltzmann equation \cite{GS03}.

In the perturbed USF problem, Eqs.\ \eqref{3.28}--\eqref{3.30} still apply with the replacements
\begin{equation}
\label{4.2}
{\cal L}X\to \nu \beta X-\frac{\zeta^{(0)}}{2}
\frac{\partial}{\partial {\bf c}}\cdot \left({\bf c}X\right),
\end{equation}
in the case of $X_{n,i}$ and $X_{T,i}$ and
\begin{equation}
\label{4.3}
{\cal L}X_{ij}\to \nu \beta X_{ij}-\frac{\zeta^{(0)}}{2}
\frac{\partial}{\partial {\bf c}}\cdot \left({\bf c}X_{ij}\right)
-\frac{\zeta_{u,ij}}{2}\frac{\partial}{\partial {\bf c}}\cdot
\left({\bf c}f^{(0)}\right),
\end{equation}
in the case of $X_{u,ij}$. In the above equations, $\zeta^{(0)}$
is the zeroth-order approximation to $\zeta$ which is given by
Eq.\ (\ref{2.25}). With the changes (\ref{4.2}) and
(\ref{4.3}) all the generalized transport coefficients can be
easily evaluated from Eqs.\ \eqref{3.28}--\eqref{3.30}. Details of these calculations are given in Appendix
\ref{appC}.

\section{Shear-rate dependence of the generalized transport coefficients}
\label{sec6}

\begin{figure}
\includegraphics[width=0.7 \columnwidth,angle=0]{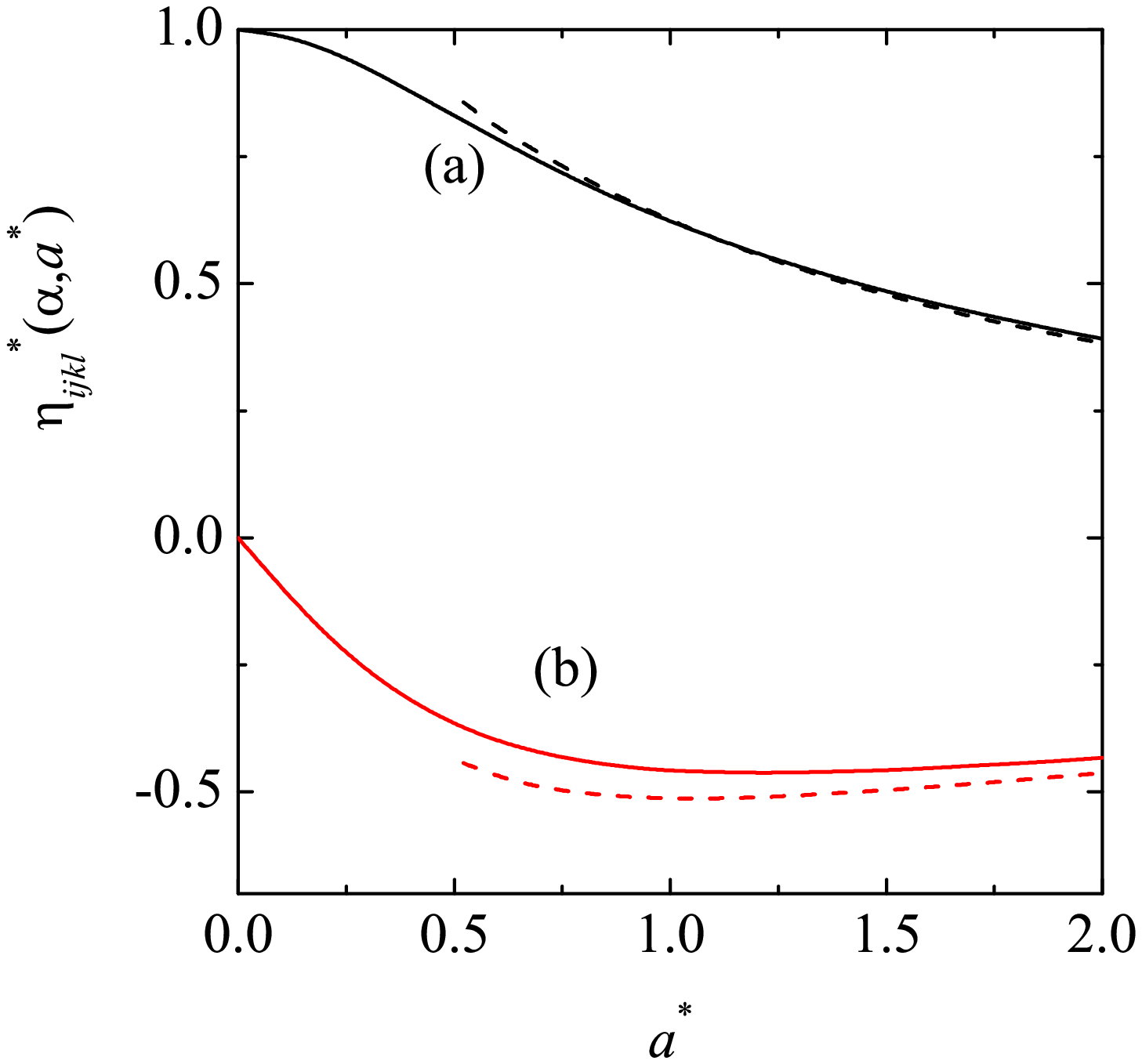}
\caption{Shear-rate dependence of the (reduced) generalized transport coefficients $\eta_{xzxz}^*$ (a) and $\eta_{yzzx}^*$ (b) for a three-dimensional ($d=3$) granular suspension with two different values of the coefficient of restitution $\al$: $\al=1$ (solid lines) and $\al=0.8$ (dashed lines).
\label{fig5.1}}
\end{figure}
\begin{figure}
\includegraphics[width=0.7 \columnwidth,angle=0]{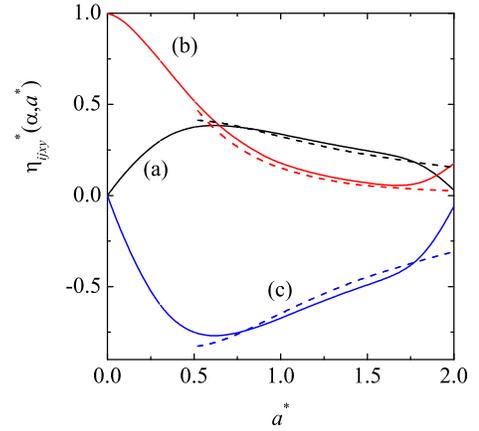}
\caption{Shear-rate dependence of the (reduced) generalized transport coefficients $\eta_{yyxy}^*$ (a), $\eta_{xyxy}^*$ (b), and  $\eta_{xxxy}^*$ (c) for a three-dimensional ($d=3$) granular suspension with two different values of the coefficient of restitution $\al$: $\al=1$ (solid lines) and $\al=0.8$ (dashed lines).
\label{fig5}}
\end{figure}

The general results derived in the previous sections clearly show that the dependence of the generalized transport coefficients on both $a^*$ and $\al$ is quite complex. Since the main goal of the present paper is to assess the shear-rate dependence of $\eta_{ijk\ell}$, $\kappa_{ij}$, and $\mu_{ij}$ for given values of $\al$, we illustrate here this dependence for some relevant elements of the above tensors by two different values of $\al$: $\al=1$ (ordinary suspensions) and $\al=0.8$ (granular suspensions). Moreover, a three-dimensional system ($d=3$) is considered in all the plots and hence, $a_\text{th}^*\simeq 0.512$ for $\al=0.8$.

To analyze the shear-rate dependence of the transport coefficients, it is convenient first to introduce the dimensionless coefficients $\eta_{ijk\ell}^*\equiv \eta_{ijk\ell}/\eta_0$, $\kappa_{ij}^*\equiv \kappa_{ij}/\kappa_0$, and
$\mu_{ij}^*\equiv n \mu_{ij}/T\kappa_0$. Here, $\eta_0=p/\nu$ and $\kappa_0=((d+2)/2)n T/(m \nu)$ are the elastic values of the shear viscosity and thermal conductivity coefficients, respectively, for a dilute gas given by the BGK kinetic model.

\subsection{Viscosity tensor}

The (reduced) elements of the viscosity tensor $\eta_{ijk\ell}^*$ are determined by solving the set of algebraic equations \eqref{c17}. There are in principle two classes of terms \cite{GS11}. Class I is made of those coefficients $\eta_{ijk\ell}^*$ with $(k,\ell)=\left\{(xx), (xy), (yx), (yy), (zz)\right\}$. The complementary class II is constituted by coefficients with $(k,\ell)=\left\{(xz), (yz), (zx), (zy)\right\}$. Of course, class II (as well as the elements $\eta_{ijzz}^*$ of class I) is meaningless in the two-dimensional case ($d=2$).

A careful analysis of the set of algebraic equations shows that the coefficients of the form $\eta_{xzk\ell}^*$ and $\eta_{yzk\ell}^*$ vanish in class I. In addition, the coefficients of the form $\eta_{xxk\ell}^*$ of class II include the first-order contribution to the cooling rate $\zeta_{u,ij}$. However, they obey a set of \emph{homogeneous} algebraic equations whose solution is the trivial one for arbitrary values of $a^*$. A similar behavior is expected for the coefficients of the form $\eta_{xyk\ell}^*$, $\eta_{yyk\ell}^*$, and $\eta_{zzk\ell}^*$. Thus, one can conclude that all the above elements of class II vanish.

The remaining elements of class II are independent of the derivatives $\partial_{a^*}P_{ij}^*$ and $\partial_{\gamma^*}P_{ij}^*$. Some of them are given by
\beq
\label{7.1}
\eta_{xzxz}^*=\eta_{yzyz}^*=\eta_{yzzy}=\frac{1+2\chi}{1-\widetilde{\gamma}+2\chi}\eta^*, \quad \eta_{yzxz}^*=0,
\eeq
\beq
\label{7.2}
\eta_{yzzx}=\frac{1+2\chi}{1-\widetilde{\gamma}+2\chi}\frac{P_{xy}^*}{P_{yy}^*}\eta^*,
\eeq
where the nonzero elements of the pressure tensor $P_{yy}^*$ and $P_{xy}^*$ are defined by Eqs.\ \eqref{2.27} and \eqref{2.28}, respectively, and the nonlinear shear viscosity $\eta^*$ is defined by Eq.\ \eqref{2.33}. The expressions of the remaining elements of class II can be obtained from Eqs.\  \eqref{c14} and \eqref{c17}. Their forms are very long and will be omitted here. Figure \ref{fig5.1} shows the dependence of two elements of class I ($\eta_{xzxz}^*$ and $\eta_{yzzx}^*$) for $\al=1$ and 0.8. These two coefficients measure the presence of non-zero values of $P_{xz}$ and $P_{yz}$ due to perturbations of the form $\partial \delta U_x/\partial z$ and $\partial \delta U_z/\partial x$, respectively. It is quite apparent that, at  a given value of $\al$, the largest impact of the shear rate on momentum transport occurs on $P_{xz}$. We also observe that $\eta_{xzxz}^*$ exhibits a shear-thinning effect more pronounced than that of the nonlinear shear viscosity $\eta^*$, as expected from Eq.\ \eqref{7.1}. In addition, the influence of collisional dissipation is very tiny in both generalized transport coefficients.

Finally, the expressions for the non-zero elements of class I contain the derivatives $\partial_{a^*}P_{ij}^*$ and $\partial_{\gamma^*}P_{ij}^*$. Those expressions are much more involved than those of class II. In order to illustrate their shear-rate dependence, we consider here the set of coefficients $\left\{\eta_{xxxy}^*, \eta_{xyxy}^*, \eta_{yyxy}^*, \eta_{zzxy}^*\right\}$. Note that $\eta_{xxxy}^*=-(\eta_{yyxy}^*+\eta_{zzxy}^*)$. In addition, the algebraic equations defining those coefficients show that $\eta_{yyxy}^*=\eta_{zzxy}^*$. This result is a consequence of the linear version of Grad's moment method that yields $P_{yy}^*=P_{zz}^*$. As said before, recent Monte Carlo simulations of granular suspensions \cite{ChVG15} have shown that the second normal stress difference is different from zero although its value is very small. The shear-rate dependence of the elements $\eta_{ijxy}^*$ is plotted in Fig.\ \ref{fig5}. The coefficients $\eta_{xyxy}^*$ and $\eta_{yyxy}^*$ measure the deviations of $P_{xy}$ and $P_{yy}$, respectively, from their \emph{unperturbed} USF values due to perturbations of the form $\partial \delta U_x/\partial y$. While the coefficient $\eta_{xyxy}^*$ decreases in general with $a^*$ (except for high shear rates), the coefficient $\eta_{yyxy}^*$ exhibits clearly a non-monotonic shear-rate dependence regardless the value of the coefficient of restitution. It is also interesting to note that $\eta_{xxxy}^*$ is always negative.
\begin{figure}
\includegraphics[width=0.7 \columnwidth,angle=0]{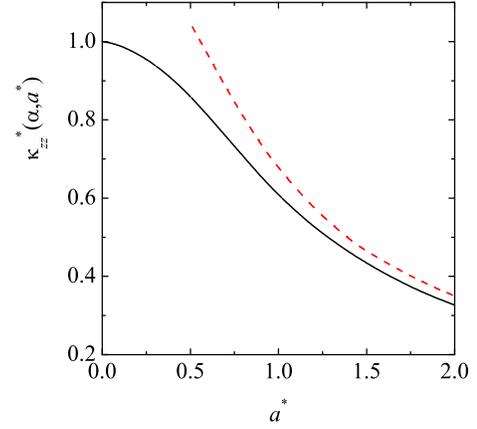}
\caption{Shear-rate dependence of the (reduced) diagonal element $\kappa_{zz}^*$ of the thermal conductivity tensor for a three-dimensional ($d=3$) granular suspension with two different values of the coefficient of restitution $\al$: $\al=1$ (solid line) and $\al=0.8$ (dashed line).
\label{fig3}}
\end{figure}
\begin{figure}
\includegraphics[width=0.7 \columnwidth,angle=0]{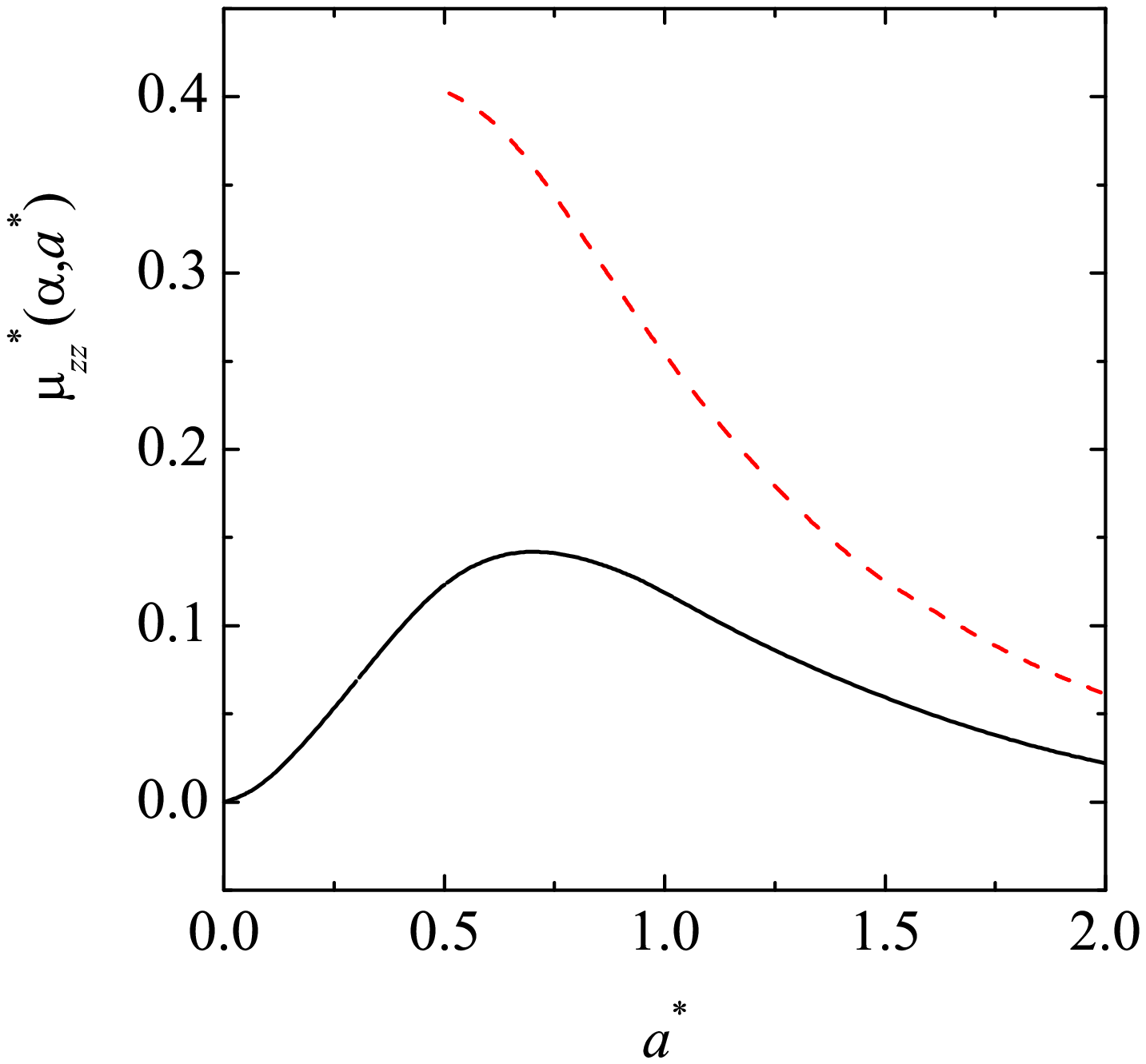}
\caption{Shear-rate dependence of the (reduced) diagonal element $\mu_{zz}^*$ of the Dufour-like tensor for a three-dimensional ($d=3$) granular suspension with two different values of the coefficient of restitution $\al$: $\al=1$ (solid line) and $\al=0.8$ (dashed line).
\label{fig4}}
\end{figure}
\begin{figure}
\includegraphics[width=0.7 \columnwidth,angle=0]{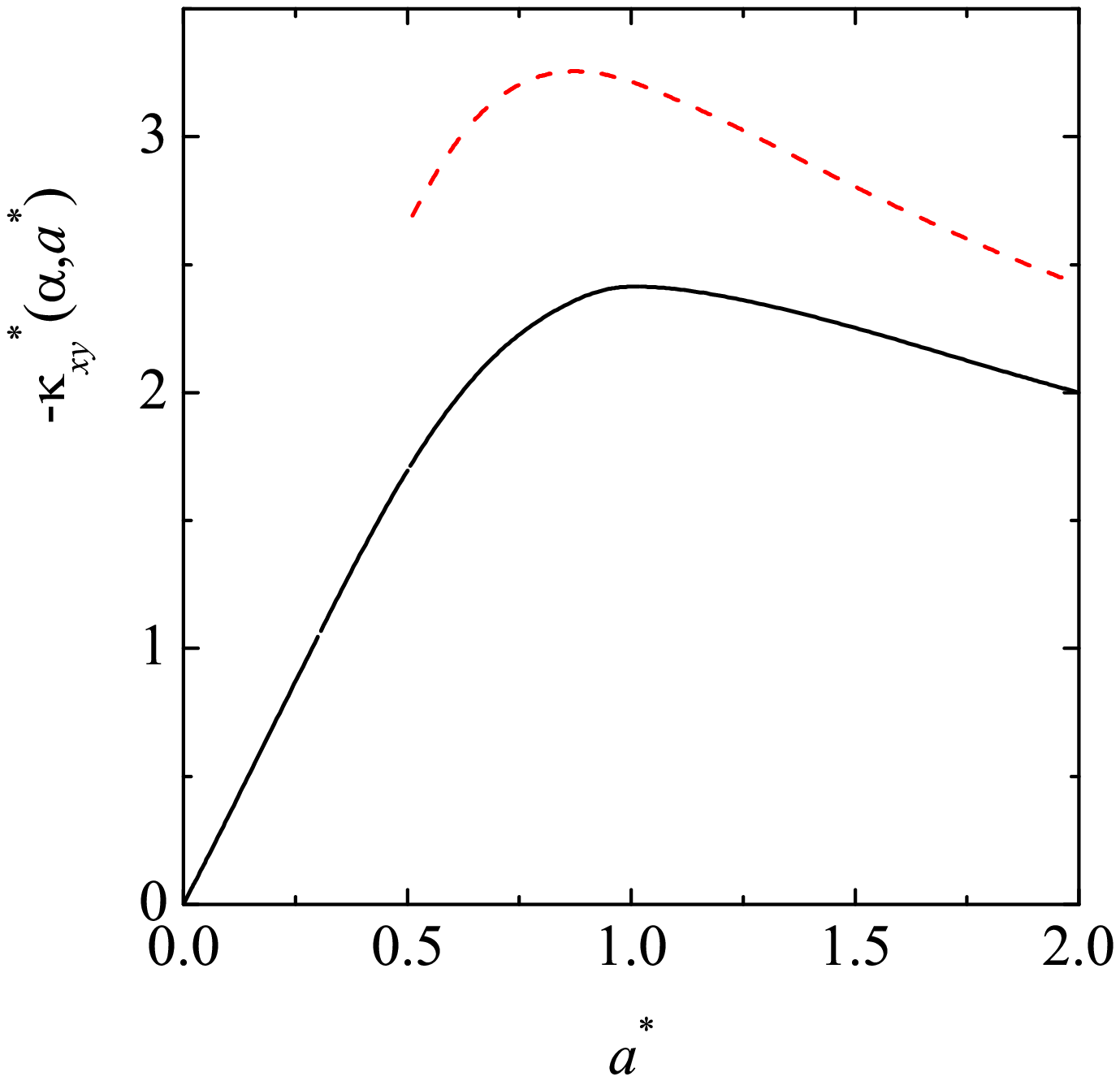}
\caption{Shear-rate dependence of the (reduced) off-diagonal element $-\kappa_{xy}^*$ of the thermal conductivity tensor for a three-dimensional ($d=3$) granular suspension with two different values of the coefficient of restitution $\al$: $\al=1$ (solid line) and $\al=0.8$ (dashed line).
\label{fig6}}
\end{figure}
\begin{figure}
\includegraphics[width=0.7 \columnwidth,angle=0]{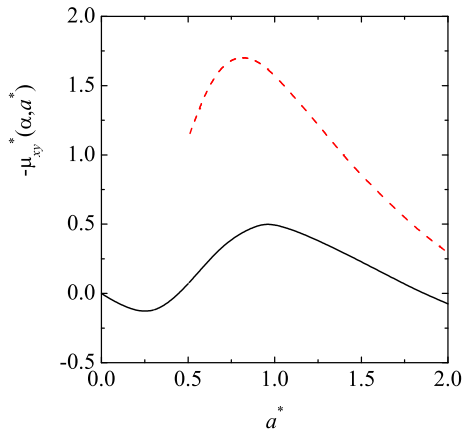}
\caption{Shear-rate dependence of the (reduced) off-diagonal element $-\mu_{xy}^*$ of the Dufour-like tensor for a three-dimensional ($d=3$) granular suspension with two different values of the coefficient of restitution $\al$: $\al=1$ (solid line) and $\al=0.8$ (dashed line).
\label{fig7}}
\end{figure}

\subsection{Thermal conductivity and Dufour-like tensors}

The evaluation of the heat flux transport coefficients $\Delta_{ij}\equiv \left\{\kappa_{ij}^*, \mu_{ij}^*\right\}$ is much more involved than that of the shear viscosity tensor $\eta_{ijk\ell}^*$. As Eqs.\ \eqref{c12} and \eqref{c13} show, in the steady state the set of transport coefficients $\Delta_{ij}$ also depends on the derivatives of the fourth-degree moments of the USF with respect to $\gamma^*$ and $a^*$. The evaluation of these derivatives is in general a quite tedious task that can be accomplished by following the steps devised in the Appendix \ref{appA}  \cite{code}.

As mentioned before, we have $\Delta_{xz}=\Delta_{zx}=\Delta_{yz}=\Delta_{zy}=0$ according to the linear shear flow \eqref{2.15}. Therefore, there are five nonzero elements of the (scaled) tensors $\Delta_{ij}$: the three diagonal ($\Delta_{xx}$, $\Delta_{yy}$, and $\Delta_{zz}$) and the two off-diagonal elements ($\Delta_{xy}$ and  $\Delta_{yx}$). The algebraic equations \eqref{c15} and \eqref{c16} also show that the anisotropy induced by the shear flow yields the properties $\Delta_{xx}\neq \Delta_{yy}\neq \Delta_{zz}$ and $\Delta_{xy} \neq \Delta_{yx}$.

To illustrate the shear-rate dependence of the coefficients $\Delta_{ij}$, we consider here the elements $\Delta_{zz}\equiv \left\{\kappa_{zz}^*, \mu_{zz}^*\right\}$ and $\Delta_{xy}\equiv \left\{\kappa_{xy}^*, \mu_{xy}^*\right\}$. The first set of coefficients measures the heat flux along the direction orthogonal to the shearing plane. The second set of coefficients provides information on cross-effects in the thermal conduction since $\kappa_{xy}^*$ and $\mu_{xy}^*$ measure the transport of energy parallel to the flow direction due to a thermal gradient along the velocity gradient. Figures \ref{fig3}--\ref{fig7} show the generalized coefficients $\kappa_{yy}^*$, $\mu_{yy}^*$, $\kappa_{xy}^*$, and $\mu_{xy}^*$ versus $a^*$ for $\al=1$ and 0.8.  We observe first that  the deviations of these coefficients with respect to their equilibrium values is significant, regardless of the collisional dissipation. This means that the impact of shear flow on heat transport is in general significant in a region of shear rates where shear thinning is quite important (see Fig.\ \ref{fig2}). Regarding the diagonal element $\kappa_{yy}^*$, it is quite apparent from Fig.\ \ref{fig3} that this coefficient decreases with $a^*$ in the region of shear rates considered. A similar behavior is found in Fig.\ \ref{fig4} for $\mu_{zz}^*$ when the collisions are inelastic ($\al \neq 1$). On the other hand, for elastic collisions, $\mu_{zz}^*$ first increases with $a^*$ for small shear rates and then it decreases with the shear rate. In any case, for elastic collisions, the magnitude of $\mu_{zz}^*$  is much smaller than that of $\kappa_{zz}^*$. Thus, for practical purposes, one can neglect the contribution to the heat flux coming from the term proportional to the density gradient when the collisions are elastic. In accordance with the above results, we conclude that in general the shear flow inhibits the transport of energy along the direction orthogonal to the velocity gradient (vorticity direction). With respect to the influence of $\al$ on both generalized coefficients, it appears that the effect of inelasticity is more important in the case of $\mu_{zz}^*$ than in the case of $\kappa_{zz}^*$. 

The absolute values of the off-diagonal elements $\kappa_{xy}^*$ and $\mu_{xy}^*$ are plotted in Figs.\ \ref{fig6} and \ref{fig7}, respectively. As said before, these coefficients measure cross effects in the energy transport. This cross coupling does not appear in the linear regime since the first-order contribution to the heat flux $q_x^{(1)}$ is at least of Burnett order (i.e., proportional to $a^* \partial_x T$). It is quite apparent that the element $\kappa_{xy}$ is negative and its magnitude presents a non-monotonic dependence with $a^*$ since it increases first with the shear rate (in the region of small shear rates), reaches a maximum and then decreases with increasing $a^*$. This behavior is much more evident in the case of elastic collisions. Regarding the coefficient $\mu_{xy}^*$, we observe that it is always negative for granular suspensions ($\al \neq 1$) and its magnitude is very small for elastic collisions. Recall that the coefficient $\mu_{ij}^*$ vanishes when $\al=1$ for vanishing shear rates. As in the case of the diagonal elements, the effect of inelasticity on heat transport is more noticeable for $\mu_{xy}^*$ than for $\kappa_{xy}^*$.

Finally, it is important to remark that the qualitative shear-rate dependence of $\kappa_{zz}^*$ and $\kappa_{xy}^*$ obtained here for elastic collisions (ordinary fluids) agrees with the one observed years ago by Daivis and Evans \cite{DE93} in molecular dynamics simulations of a thermostatted shear-flow state.

\section{Concluding remarks}
\label{sec7}

The influence of gas phase on the transport properties of solid particles under USF has been studied in this paper. In the low-density regime, a viscous drag force term for the interstitial fluid has been incorporated into the Boltzmann kinetic equation to account for the effect of the former on the dynamics of grains. The physical situation is such that the granular suspension is in a state that deviates from the USF by small spatial gradients. Since the system is subjected to a strong shear flow and is not restricted to nearly elastic spheres, the corresponding transport coefficients characterizing momentum and heat transport are nonlinear functions of both the shear rate and the coefficient of restitution. The explicit determination of the above coefficients has been the main objective of the present contribution. The search for such expressions has been prompted by previous results \cite{L06,G06} obtained for \emph{dry} granular gases (i.e., in the absence of the viscous drag force). Here, the problem is revisited by considering the effect of the gas phase on transport properties.

Assuming that the USF state is slightly perturbed, the Boltzmann equation \eqref{2.4} has been solved by means of a Chapman-Enskog-like expansion. The new feature of this expansion is that the (local) shear flow distribution is employed as the reference state instead of the usual (local) equilibrium distribution \cite{CC70} or the (local) homogeneous cooling state \cite{BDKS98,GD99}. As already noted in previous works \cite{L06,G06,GMT13,GChV13}, since the zeroth-order derivative $\partial_t^{(0)} T$ is in general different from zero, the reference base state is not stationary. This fact introduces technical difficulties in the implementation of the perturbation scheme. Thus, in order to get explicit results, the steady-state condition \eqref{2.17} is considered at the end of the calculations. In this state, the (reduced) shear rate $a^*$ and the coefficient of restitution $\al$ are coupled to the (scaled) friction coefficient $\gamma^*$, so that the former two are the relevant parameters of the problem.

To first order of the expansion, the momentum and heat fluxes are given by Eqs.\ \eqref{3.22} and \eqref{3.23}, respectively, where the generalized transport coefficients $\eta_{ijk\ell}$, $\kappa_{ij}$, and $\mu_{ij}$ are defined in terms of the solutions of the set of coupled integral equations\eqref{3.28}--\eqref{3.30}. However, since the solution of the above integral equations is in general quite a complex problem, the BGK-like kinetic model \eqref{4.1} has been employed to obtain the explicit shear-rate dependence of the above set of transport coefficients. Although the kinetic model \eqref{4.1} can be considered as a crude representation of the true Boltzmann equation, it gives the same results for the rheological properties as those derived from the Boltzmann equation  by means of Grad's moment method. Given that those theoretical predictions compare quite well with Monte Carlo simulations \cite{ChVG15}, it is expected that the results provided by the kinetic model are accurate even for conditions of practical interest, such as strong dissipation and/or large shear rates.

As expected, there are many new transport coefficients in comparison to the case of states close to equilibrium (for ordinary gases) or states near the homogeneous cooling state (for dry granular gases). Here, for the sake of illustration, the shear-rate dependence of some relevant elements of the viscosity tensor, the thermal conductivity tensor, and the Dufour-like tensor have been studied. More specifically,  Figs.\ \ref{fig5.1} and \ref{fig5} show the (reduced) elements $\eta_{xzxz}^*$, $\eta_{yzzx}^*$, and $\eta_{ijxy}^*$, respectively, Figs.\ \ref{fig3} and \ref{fig4} show the diagonal elements $\kappa_{zz}^*$ and $\mu_{zz}^*$, respectively, and Figs.\ \ref{fig6} and \ref{fig7} show the off-diagonal elements $\kappa_{xy}^*$ and $\mu_{xy}^*$, respectively. It is apparent that in general the deviation of these coefficients from their equilibrium values (i.e., for $a^*=0$ and $\al=1$) is quite significant. In addition, the influence of collisional dissipation on transport is much more significant for the heat flux transport coefficients than for the coefficients associated with the pressure tensor.

\begin{figure}
\includegraphics[width=0.7 \columnwidth,angle=0]{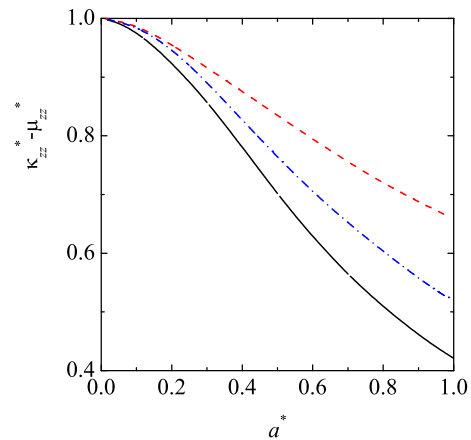}
\caption{Shear-rate dependence of the (reduced) element $\widehat{\kappa}_{zz}\equiv \kappa_{zz}^*-\mu_{zz}^*$ for a three-dimensional ($d=3$) ordinary fluid ($\al=1$). The solid line corresponds to the results obtained here, the dashed line refers to the results derived from the Boltzmann equation in Ref.\ \cite{G95} for Maxwell molecules, and the dash--dotted line corresponds to the results obtained in Ref.\ \cite{G93} from the BGK equation for Maxwell molecules.
\label{fig8}}
\end{figure}

As said in the Introduction, for ordinary fluids ($\al=1$), the thermal conductivity tensor of a thermostatted shear-flow state was determined years ago from the BGK \cite{G93} and Boltzmann \cite{G95} kinetic equations. The physical situation corresponds to a perturbed \emph{steady} USF state with $\delta \mathbf{U}=\mathbf{0}$, $p=n T\equiv \text{const.}$ and $\nabla T \neq 0$. Under these conditions, one needs to add an external field that exactly compensates for the increase or decrease of momentum due to the term $\nabla \cdot \mathsf{P}$ \cite{GS03}. The addition of this external field affects the value of the thermal conductivity tensor and hence, the situation studied in Refs.\ \cite{G93,G95} slightly differs from the one analyzed in the present paper. On the other hand, in order to make a comparison with these previous results \cite{G93,G95}, one considers particular perturbations such that $\nabla p=0$ and so, $\nabla \ln n=-\nabla \ln T$. Therefore,
the heat flux \eqref{3.23} obeys the generalized Fourier's law
\beq
\label{6.1}
q_i^{(1)}=-\kappa_0 \widetilde{\kappa}_{ij} \partial_j T, \quad \widetilde{\kappa}_{ij}=\kappa_{ij}^*-\mu_{ij}^*.
\eeq
Figure \ref{fig8} shows the transport coefficient $\widetilde{\kappa}_{zz}\equiv \kappa_{zz}^*-\mu_{zz}^*$ versus $a^*$ for $\al=1$. We observe that the previous predictions made for thermostatted shear flow states from the BGK \cite{G93} and Boltzmann \cite{G95} equations compare qualitatively well with the results obtained here for arbitrary perturbations. However, at a more quantitative level, it seems that the impact of shear flow on energy transport is more significant in the situation analyzed in this paper than those studied in Refs.\ \cite{G93,G95}.

\begin{figure}
\includegraphics[width=0.7 \columnwidth,angle=0]{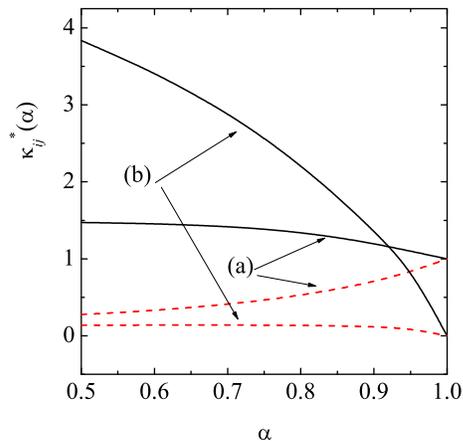}
\caption{Plot of the (reduced) elements $\kappa_{yy}^*$ (a) and $-\kappa_{xy}^*$ (b) as a function of the coefficient of restitution $\al$ for a two-dimensional dry granular gas ($\gamma^*=0$). The solid and dashed lines are the results derived in this paper and in Ref.\ \cite{SA14}, respectively.
\label{fig9}}
\end{figure}

In the case of dry granular gases ($\gamma^*=0$ but $\al\neq 1$), Saha and Alam \cite{SA14} have determined the heat flux of a two-dimensional granular gas under USF. The results were obtained by solving the Boltzmann equation by means of a perturbation expansion around an anisotropic Gaussian distribution. This distribution was employed years ago by Jenkins and Richman \cite{JR88} to obtain the rheological properties of USF via Grad's moment method. The corresponding constitutive relation for the heat flux derived in Ref.\ \cite{SA14} can be written as
\beq
\label{6.2}
q_i^{(1)}=-\kappa_{ij} \partial_j T-\Xi_{ij}\partial_j \Pi_{ij},
\eeq
where $\Pi_{ij}$ is the deviatoric or traceless part of the pressure tensor defined by Eq.\ \eqref{2.22}. In Eq.\ \eqref{6.2}, $\kappa_{ij}$ is identified as the thermal conductivity tensor and $\Xi_{ij}$ is a tensor quantifying the contribution to the heat flux coming from the gradient of the deviatoric stress $\Pi_{ij}$. As expected, the tensors $\kappa_{ij}$ and $\Xi_{ij}$ are nonlinear functions of the coefficient of restitution $\al$. It appears first that Eq.\ \eqref{6.2} disagrees with the constitutive relation \eqref{3.23} derived here for the heat flux. On the other hand, in an attempt to make a comparison with the theoretical results obtained in Ref.\ \cite{SA14} for the thermal conductivity tensor, Fig.\ \ref{fig9} shows the dependence of the (reduced) coefficients $\kappa_{yy}^*$ and $\kappa_{xy}^*$ on $\al$ for a dry two-dimensional granular gas. Notice that, in order to get analytical expressions, the theoretical results of Ref.\ \cite{SA14} plotted in Fig.\ \ref{fig9} were derived by considering terms up to super-Burnett order (i.e., third order in the shear rate). We observe that the $\al$-dependence of the diagonal element $\kappa_{yy}^*$ is qualitatively different from the one predicted in Ref.\ \cite{SA14} since while in the latter theory $\kappa_{yy}^*$ decreases with increasing inelasticity, the opposite happens here. Although a more qualitative agreement is found for the magnitude of $\kappa_{xy}^*$, both theoretical results exhibit significant quantitative discrepancies for strong inelasticity. The differences between both theories at the level of the thermal conductivity tensor $\kappa_{ij}$ could be in part due to the different form of the constitutive relation for the heat flux derived in Ref.\ \cite{SA14}. In addition, while the results obtained in the latter work were obtained by solving the Boltzmann equation up to super-Burnett order, the theoretical predictions made in the present paper are based on an exact solution of the BGK-like kinetic model. It would be convenient to perform computer simulations for $\kappa_{ij}$ to check the reliability of the above theories for strong inelasticities.

The explicit results reported in this paper can be useful for studying different problems First, as done in Ref.\ \cite{G06}, an important application is to perform an stability analysis of the hydrodynamic equations with respect to the USF state. This analysis will allow us to identify the conditions for stability in terms of both the shear rate and the coefficient of restitution. Another interesting and challenging problem is to extend the present results by considering the general Langevin-like model proposed in Ref.\ \cite{GTSH12}. This will allow us to provide additional refinements of the predictions obtained here so that, a closer comparison with direct numerical simulations of granular suspensions could be performed. Finally, it would be also relevant to extend the analysis made here for a monodisperse granular suspension to the intriguing and important subject of polydisperse suspensions. A good starting point for this achievement could be the suspension model introduced in Ref.\ \cite{KG13}. Work along the above lines will be carried out in the near future.

\acknowledgments

I am grateful to Dr. Mariano L\'opez de Haro for a critical reading of the manuscript. The present research has been supported by the Spanish Government through grant No. FIS2016-76359-P, partially financed by FEDER funds and by the Junta de Extremadura (Spain) through Grant No. GR15104.

\appendix
\section{Derivatives of the zeroth-order velocity moments with respect to $\gamma^*$ and $a^*$ in the steady state}
\label{appA}

The derivatives of the zeroth-order velocity moments with respect to $\gamma^*$ and $a^*$ in the steady state are determined in this appendix. We start with the pressure tensor $P_{ij}^{(0)}$, whose elements obey Eq.\ \eqref{3.13}. In dimensionless form, Eq.\ \eqref{3.13} is given by
\beqa
\label{a1}
& & -\left(\frac{2}{d}a^*
P_{xy}^{*}+2\gamma^*+\zeta_0^*\right)\left[P_{ij}^{*}-\frac{1}{2}\left(\gamma^* P_{ij,\gamma}^*
+a^*P_{ij,a}^*\right)\right]\nonumber\\
& & +a_{ik}^*P_{kj}^*+a_{jk}^*P_{ki}^*+2\gamma^*P_{ij}^*=\beta \delta_{ij}-\left(\beta+\zeta_0^*\right)P_{ij}^*,\nonumber\\
\eeqa
where
\beq
\label{a1.1}
P_{ij,\gamma}^*\equiv \frac{\partial P_{ij}^*}{\partial\gamma^*}, \quad
P_{ij,a}^*\equiv \frac{\partial P_{ij}^*}{\partial a^*},
\eeq
and upon deriving Eq.\ \eqref{a1} use has been made of the relation \eqref{3.16}.  Let us consider the elements $P_{yy}^*=P_{zz}^*$ and $P_{xy}^*$. From Eq.\ \eqref{a1}, one gets
\beqa
\label{a3}
& & -\left(\frac{2}{d}a^*
P_{xy}^{*}+2\gamma^*+\zeta_0^*\right)\left[P_{yy}^{*}-\frac{1}{2}\left(\gamma^* P_{yy,\gamma}^*
\right.\right.\nonumber\\
& & \left.\left.+a^*P_{yy,a}^*\right)\right]+2\gamma^*P_{yy}^*=\beta -\left(\beta+\zeta_0^*\right)P_{yy}^*,
\eeqa
\beqa
\label{a4}
& & -\left(\frac{2}{d}a^*
P_{xy}^{*}+2\gamma^*+\zeta_0^*\right)\left[P_{xy}^{*}-\frac{1}{2}\left(\gamma^* P_{xy,\gamma}^*
\right.\right.\nonumber\\
& & \left.\left.
+a^*P_{xy,a}^*\right)\right]+a^* P_{yy}^*+2\gamma^*P_{xy}^*= -\left(\beta+\zeta_0^*\right)P_{xy}^*.
\nonumber\\
\eeqa
The goal here is to evaluate the derivatives $P_{yy,\gamma}^*$, $P_{yy,a}^*$, $P_{xy,\gamma}^*$ and $P_{xy,a}^*$ at the steady state. This state is defined by the condition \eqref{2.17}. To get these derivatives, we differentiate first Eqs.\ \eqref{a3} and \eqref{a4} with respect to $a^*$ and take then the steady-state limit. The result is
\beqa
\label{a5}
& &-\frac{2}{d}\left(P_{xy}^{*}+a^* P_{xy,a}^*\right)\left[P_{yy}^{*}-\frac{1}{2}\left(\gamma^* P_{yy,\gamma}^*
+a^*P_{yy,a}^*\right)\right]\nonumber\\
& & +2\gamma^*P_{yy,a}^*=-\left(\beta+\zeta_0^*\right)P_{yy,a}^*,
\eeqa
\beqa
\label{a6}
& & -\frac{2}{d}\left(P_{xy}^{*}+a^* P_{xy,a}^*\right)\left[P_{xy}^{*}-\frac{1}{2}\left(\gamma^* P_{xy,\gamma}^*
+a^*P_{xy,a}^*\right)\right]\nonumber\\
& & +P_{yy}^*+a^* P_{yy,a}^*+2\gamma^*P_{xy,a}^*=-\left(\beta+\zeta_0^*\right)P_{xy,a}^*,
\eeqa
where here it is understood that all the terms are evaluated at the steady state. To close the problem, we differentiate then Eqs.\ \eqref{a3} and \eqref{a4} with respect to $\gamma^*$ and take the steady-state limit with the result
\beqa
\label{a7}
& & -\frac{2}{d}\left(d+a^* P_{xy,\gamma}^*\right)\left[P_{yy}^{*}-\frac{1}{2}\left(\gamma^* P_{yy,\gamma}^*
+a^*P_{yy,a}^*\right)\right]\nonumber\\
& & +2\left(P_{yy}^*+\gamma^*P_{yy,\gamma}^*\right)=-\left(\beta+\zeta_0^*\right)P_{yy,\gamma}^*,
\eeqa
\beqa
\label{a8}
& & -\frac{2}{d}\left(d+a^*P_{xy,\gamma}^{*}\right)\left[P_{xy}^{*}-\frac{1}{2}\left(\gamma^* P_{xy,\gamma}^*
+a^*P_{xy,a}^*\right)\right]\nonumber\\
& & +a^* P_{yy,\gamma}^*+2\left(P_{xy}^*+\gamma^*P_{xy,\gamma}^*\right)=-\left(\beta+\zeta_0^*\right)P_{xy,\gamma}^*.
\nonumber\\
\eeqa
The set of nonlinear algebraic equations \eqref{a5}--\eqref{a8} can be numerically solved for given values of $a^*$ and $\al$. In the dry limit case ($\gamma^*=0$), the solution to Eqs.\ \eqref{a5} and \eqref{a6} can be written as
\beq
\label{a9}
P_{yy,a}^*=4 P_{yy}^*\frac{a^*P_{xy,a}^*+P_{xy}^*}{2d\beta+d\zeta_0^*+2a^{*2}P_{xy,a}^*},
\eeq
where $P_{xy,a}^*$ is the real root of the cubic equation
\beqa
\label{a10}
& &2a^{*4}P_{xy,a}^{*3}+4da^{*2}(\zeta_0^*+\beta)P_{xy,a}^{*2}+\frac{d^2}{2}\left(7\zeta_0^*+14\zeta_0^*\beta\right.
\nonumber\\
& &\left.+4\beta^2\right)P_{xy,a}^*+d^2\beta (\zeta_0^*+\beta)^{-2}\left(2\beta^2-2\zeta_0^{*2}-\beta \zeta_0^*\right)=0.
\nonumber\\
\eeqa
Equations \eqref{a9} and \eqref{a10} agree with previous results \cite{L06,G06} derived for a dry granular gas of inelastic hard spheres.

The corresponding derivatives of the fourth-degree velocity moments of the distribution $f^{(0)}$ with respect to $\gamma^*$ and $a^*$ in the steady state are also needed to determine the generalized coefficients $\kappa_{ij}$ and $\mu_{ij}$ associated with the first-order contribution to the heat flux. To evaluate these derivatives, the BGK kinetic model \eqref{4.1} is considered. The velocity moments of the distribution $f^{(0)}$ are defined as
\begin{equation}
\label{a11}
M_{k_1,k_2,k_3}^{(0)}=\int \; \dd {\bf c}\;
c_x^{k_1}c_y^{k_2}c_z^{k_3} f^{(0)}({\bf c}).
\end{equation}
These moments verify the equation
\beqa
\label{a12}
& & -\left(\frac{2}{d}\widetilde{a}
P_{xy}^{*}+2\widetilde{\gamma}+\widetilde{\zeta}_0\right)T\partial_T M_{k_1,k_2,k_3}^{(0)}+\widetilde{a}k_1
\nonumber\\
& & \times M_{k_1-1,k_2+1,k_3} +\left(1+k\widetilde{\xi}\right)M_{k_1,k_2,k_3}=N_{k_1,k_2,k_3},
\nonumber\\
\eeqa
where $k\equiv k_1+k_2+k_3$, $\widetilde{a}\equiv a^*/\beta$, $\widetilde{\zeta}_0\equiv \zeta_0^*/\beta$, $\widetilde{\gamma}\equiv \gamma^*/\beta$, $\widetilde{\xi}=\widetilde{\gamma}+\widetilde{\zeta}_0/2$, and $N_{k_1,k_2,k_3}$ are the velocity moments of $f_\text{M}$. In the steady state, $\widetilde{\xi}=\chi$ where $\chi$ is given by Eq.\ \eqref{2.30}. As in the case of the pressure tensor, the
derivative $T \partial_T M_{k_1,k_2,k_3}^{(0)}$ can be written as
\beqa
\label{a14}
& &T\partial_T M_{k_1,k_2,k_3}^{(0)}=T\partial_T n\left(\frac{2T}{m}
\right)^{k/2}M_{k_1,k_2,k_3}^*(\gamma^*,a^*)\nonumber\\
&=&\frac{1}{2}n\left(\frac{2T}{m}\right)^{k/2} \left(k M_{k_1,k_2,k_3}^{*}
-\gamma^* M_{k_1,k_2,k_3,\gamma}^{*}\right.\nonumber\\
& & \left.-a^*M_{k_1,k_2,k_3,a}^{*}\right),
\nonumber\\
\eeqa
where we have introduced the shorthand notation
\beq
\label{a16}
M_{k_1,k_2,k_3,\gamma}^{*} \equiv \partial_{\gamma^*}M_{k_1,k_2,k_3}^{*},
\eeq
\beq
\label{a16.1}
M_{k_1,k_2,k_3, a}^{*} \equiv \partial_{a^*}M_{k_1,k_2,k_3}^{*},
\eeq

In dimensionless form, Eq.\ (\ref{a12}) reads
\beqa
\label{a15}
& & -\left(\frac{2}{d}\widetilde{a}
P_{xy}^{*}+2\widetilde{\gamma}+\widetilde{\zeta}_0\right)\frac{1}{2}\left(k M_{k_1,k_2,k_3}^{*}
-\gamma^*M_{k_1,k_2,k_3,\gamma}^{*}\right.\nonumber\\
& & \left.-a^*M_{k_1,k_2,k_3,a}^{*}\right)+ k_1 \widetilde{a}
M_{k_1-1,k_2+1,k_3}^{*}+
\left(1+k\widetilde{\xi}\right)\nonumber\\
& & \times M_{k_1,k_2,k_3}^{*}-N_{k_1,k_2,k_3}^{*}=0,
\eeqa
where $M_{k_1,k_2,k_3}^{*}\equiv n^{-1}(m/2T)^{k/2}M_{k_1,k_2,k_3}$, and
\begin{equation}
\label{a17}
N_{k_1,k_2,k_3}^{*}=\pi^{-3/2}\Gamma \left(\frac{k_1+1}{2}\right)
\left(\frac{k_2+1}{2}\right)\left(\frac{k_3+1}{2}\right)
\end{equation}
if $k_1$, $k_2$, and $k_3$ are even, being zero otherwise. Equation \eqref{a15} provides the expressions of the reduced moments
$M_{k_1,k_2,k_3,s}^{*}$ in the steady state (e.g., when $\frac{2}{d}\widetilde{a}
P_{xy}^{*}+2\widetilde{\gamma}+\widetilde{\zeta}_0=0$).

In order to evaluate the derivatives $M_{k_1,k_2,k_3,\gamma}^{*}$ and $M_{k_1,k_2,k_3,a}^{*}$ in the steady state, we differentiate with respect to $\gamma^*$ and $a^*$, respectively, both sides of Eq.\ \eqref{a15} and then take the steady state condition \eqref{2.17}. As an illustration, let us consider the moment $M_{040}^*$ which obeys the equation
\beqa
\label{a18}
& & -\left(\frac{2}{d}\widetilde{a}
P_{xy}^{*}+2\widetilde{\gamma}+\widetilde{\zeta}\right)\left(2-\frac{1}{2}\gamma^*\partial_{\gamma^*}
-\frac{1}{2}a^*\partial_{a^*}\right)M_{040}^{*}\nonumber\\
& & +\left(1+4\widetilde{\xi}\right)M_{040}^*=\frac{3}{4}.
\eeqa
From Eq.\ \eqref{a18}, in the steady state, one gets the identities
\beqa
\label{a19}
& & -\left(\frac{2}{d}\widetilde{a}P_{xy,\gamma}^{*}+2 \beta^{-1}\right)\left[2M_{040}^*-\frac{1}{2}\left(\gamma^*M_{040,\gamma}^{*}\right.\right.\nonumber\\
& & \left.\left.
+a^*M_{040,a}^{*}\right)\right]+4\beta^{-1}M_{040}^{*}
+\left(1+4\widetilde{\xi}\right)M_{040,\gamma^*}^{*}\nonumber\\
& & =0,
\eeqa
\beqa
\label{a20}
& & -\frac{2}{d}\left(\beta^{-1}P_{xy}^*+\widetilde{a}P_{xy,a}^{*}\right)
\left[2M_{040}^*-\frac{1}{2}\left(\gamma^*M_{040,\gamma}^{*}\right.\right.\nonumber\\
& & \left.\left.
+a^*M_{040,a}^{*}\right)\right]+\left(1+4\widetilde{\xi}\right)M_{040,a}^{*}=0.
\eeqa
The solution to the set of linear algebraic equations \eqref{a19} and \eqref{a20} gives the derivatives $M_{040,\gamma}^{*}$ and $M_{040,a}^{*}$ in terms of $a^*$ and $\al$. Proceeding in a similar way, all the derivatives of the fourth-degree velocity moments with respect to both $a^*$ and $\gamma^*$ can be analytically computed in the steady state.

\section{First-order approximation}
\label{appB}

The kinetic equation for the first-order distribution $f^{(1)}$ is
\beqa
\label{b1}
& & \partial_t^{(0)}f^{(1)}-aV_y\frac{\partial f^{(1)}}{\partial V_x}
-\gamma\frac{\partial}{\partial{\bf V}}\cdot {\bf V} f^{(1)}
+{\cal L}f^{(1)}\nonumber\\
& &=
-\left[\partial_t^{(1)}+({\bf V}+{\bf U}_0)\cdot \nabla \right]f^{(0)}.
\eeqa
The velocity dependence on the right side of Eq.\ (\ref{b1}) can be obtained from the macroscopic
balance equations to first order in the gradients. They are given by
\begin{equation}
\label{b3}
\partial_t^{(1)}n+{\bf U}_0\cdot \nabla n=-\nabla \cdot (n\delta {\bf U}),
\end{equation}
\begin{equation}
\label{b4}
\partial_t^{(1)}\delta {\bf U}+({\bf U}_0+\delta {\bf U})\cdot \nabla \delta {\bf U}=
-\frac{1}{\rho}\nabla \cdot {\sf P}^{(0)},
\end{equation}
\beqa
\label{b5}
\frac{d}{2}n\partial_t^{(1)}T&+&\frac{d}{2}n({\bf U}_0+
\delta {\bf U})\cdot \nabla T+aP_{xy}^{(1)} \nonumber\\
& & +{\sf P}^{(0)}:\nabla\delta {\bf U}=-\frac{d}{2}p\zeta^{(1)},
\eeqa
where $\rho=mn$ is the mass density,
\begin{equation}
\label{b6}
P_{ij}^{(1)}=\int\; \dd{\bf c}\, m c_i c_j  f^{(1)}({\bf c}),
\end{equation}
and
\begin{equation}
\label{b7}
\zeta^{(1)}=\frac{1}{d p}\int\; \dd{\bf c}\,mc^2{\cal L}f^{(1)}.
\end{equation}
Use of Eqs.\ \eqref{b3}--\eqref{b5} in Eq.\ \eqref{b1} yields
\beqa
\label{b8}
& &\left(\partial_t^{(0)}-aV_y\frac{\partial}{\partial V_x}
-\gamma \frac{\partial}{\partial{\bf V}}\cdot {\bf V} f^{(1)}+{\cal L}\right)
f^{(1)}\nonumber\\
& &-\zeta^{(1)}T\frac{\partial f^{(0)}}{\partial T} ={\bf Y}_n\cdot \nabla n+{\bf Y}_T\cdot \nabla T
+{\sf Y}_u:\nabla \delta {\bf U},
\nonumber\\
\eeqa
where
\begin{equation}
\label{b9}
Y_{n,i}=-\frac{\partial f^{(0)}}{\partial n}c_i-\frac{1}{\rho}
\frac{\partial f^{(0)}}{\partial c_j}\frac{\partial P_{ij}^{(0)}}{\partial n},
\end{equation}
\begin{equation}
\label{b10}
Y_{T,i}=-\frac{\partial f^{(0)}}{\partial T}c_i-\frac{1}{\rho}
\frac{\partial f^{(0)}}{\partial c_j}\frac{\partial P_{ij}^{(0)}}{\partial T},
\end{equation}
\begin{equation}
\label{b11}
Y_{u,ij}=n\frac{\partial f^{(0)}}{\partial n}\delta_{ij}+c_j\frac{\partial f^{(0)}}
{\partial c_i}+\frac{2}{d n}\frac{\partial f^{(0)}}{\partial T}\left(P_{ij}^{(0)}-a\eta_{xyij}\right).
\end{equation}

According to the symmetry properties of $f^{(1)}$, the only nonzero contribution to $\zeta^{(1)}$ comes from the term proportional to the tensor $\nabla_i \delta U_j$. Thus,
\begin{equation}
\label{b12}
\zeta^{(1)}={\zeta}_{u,ji}\nabla_{i} \delta U_j.
\end{equation}
An estimation of $\zeta_{u,ij}$ has been made in Ref.\ \cite{G06} for a three-dimensional system ($d=3$). The result is
\begin{equation}
\label{b13}
\zeta_{u,ij}=-\frac{1}{15}\sigma^2\sqrt{\frac{\pi}{m T}}(1-\alpha^2)
\Pi_{k\ell}^*\eta_{k\ell ij},
\end{equation}
where $\Pi_{ij}^*\equiv \Pi_{ij}/n T$. Of course, when $\alpha=1$, then $\zeta_{u,ij}=0$.

The solution to Eq.\ (\ref{b8}) has the form
\begin{equation}
\label{b14}
f^{(1)}=X_{n,i}({\bf c})\nabla_i n+ X_{T,i}({\bf c})\nabla_i T+X_{u,ji}({\bf c})\nabla_i \delta U_j.
\end{equation}
Note that in Eq.\ (\ref{b11}) the coefficients $\eta_{ijk\ell}$ are defined through Eq.\ (\ref{3.24}). The coefficients $X_{n,i}$, $X_{T,i}$, and $X_{u,ij}$ are functions of the peculiar velocity ${\bf c}$ and
the hydrodynamic fields. In addition, there are contributions from the time derivative $\partial_t^{(0)}$
acting on the temperature and velocity gradients given by
\beqa
\label{b15}
& &\partial_t^{(0)} \nabla_i T=\left(\frac{2a}{d n^2}(1-n\partial_n)
P_{xy}^{(0)}-\frac{\zeta^{(0)}T}{n}\right) \nabla_i n\nonumber\\
& &-\left(
\frac{2a}{d n}\partial_T P_{xy}^{(0)}+2\gamma+\frac{3}{2}\zeta^{(0)}\right)\nabla_i T,
\eeqa
\beq
\label{b16}
\partial_t^{(0)} \nabla_i \delta U_j=-a_{jk} \nabla_i \delta U_k-\gamma
\nabla_i \delta U_j.
\eeq
Substituting Eqs.\ \eqref{b13}, \eqref{b15}, and \eqref{b16} into Eq.\ \eqref{b8} and identifying coefficients
of independent gradients, one finally gets the set of coupled linear integral equations
\beqa
\label{b19}
& &-\left(\frac{2}{d p}a
P_{xy}^{(0)}+2\gamma+\zeta^{(0)}\right)T \partial_T X_{n,i}- a
c_y\frac{\partial X_{n,i}}{\partial c_x}\nonumber\\
& & -\gamma\frac{\partial}{\partial{\bf c}}\cdot {\bf c}X_{n,i}
+{\cal L}X_{n,i}=Y_{n,i}
\nonumber\\
& & -
\frac{T}{n}\left[\frac{2a}{d p}(1-n\partial_n)
P_{xy}^{(0)}-\zeta^{(0)}\right]X_{T,i},
\eeqa
\beqa
\label{b20}
& &-\left(\frac{2}{d p}a
P_{xy}^{(0)}+2\gamma+\zeta^{(0)}\right)T \partial_T X_{T,i}-a
c_y\frac{\partial}{\partial c_x}X_{T,i}\nonumber\\
& & -\gamma\frac{\partial}{\partial{\bf c}}\cdot {\bf c}X_{T,i}
-\left[\frac{2a}{d n}(\partial_T P_{xy}^{(0)})+2\gamma+\frac{3}{2}\zeta^{(0)}\right]X_{T,i}\nonumber\\
& &
+{\cal L}X_{T,i}=Y_{T,i},
\eeqa
\beqa
\label{b21}
& & -\left(\frac{2}{d p}a P_{xy}^{(0)}+2\gamma+\zeta^{(0)}\right)T\partial_T X_{u,k\ell}-a
c_y\frac{\partial}{\partial c_x}X_{u,k\ell}\nonumber\\
& &-a\delta_{ky}X_{u,x\ell}-\zeta_{u,k\ell}T\partial_T
f^{(0)}-\gamma X_{u,k\ell}-\gamma\frac{\partial}{\partial{\bf c}}\cdot {\bf c}X_{u,k\ell}
\nonumber\\
& & +{\cal L}X_{u,k\ell}=Y_{u,k\ell}.
\eeqa
Upon writing Eqs.\ (\ref{b19})--(\ref{b21}), use has been made of the property
\begin{eqnarray}
\label{b22}
\partial_t^{(0)} X &=&\frac{\partial X}{\partial T}\partial_t^{(0)}
T+\frac{\partial X}{\partial \delta U_i}\partial_t^{(0)}
\delta U_i \nonumber\\
&=&-\left(\frac{2}{d n}a
P_{xy}^{(0)}+2T \gamma+T\zeta^{(0)}\right)\frac{\partial X}{\partial T}
\nonumber\\
& &
+\left(a_{ij}\delta U_j +\gamma \delta U_i\right)\frac{\partial X}{\partial c_i},
\end{eqnarray}
where in the last step we have taken into account that $X$ depends on
$\delta {\bf U}$ through ${\bf c}={\bf V}-\delta {\bf U}$.

\section{Kinetic model results in the steady USF state}
\label{appC}


In this appendix, the steady state solution to the BGK-like kinetic model \eqref{4.1} in the steady (unperturbed) USF is briefly analyzed. In this case, $\delta \mathbf{U}=\mathbf{0}$ and so $\mathbf {c}=\mathbf{V}$. In the steady state, $f_s(\mathbf{V})$ verifies the kinetic equation
\beq
\label{c.0}
-aV_y\frac{\partial f_s}{\partial V_x}-\gamma \frac{\partial}{\partial
{\bf V}}\cdot {\bf V} f_s =-\beta\nu \left(f_s-f_\text{M}\right)+\frac{\zeta^{(0)}}{2}\frac{\partial}{\partial
{\bf V}}\cdot {\bf V} f_s,
\eeq
where here $\zeta$ has been approximated by its Maxwellian approximation $\zeta^{(0)}$ given by Eq.\ \eqref{2.25}. Let us introduce the velocity moments of $f_s$ as
\begin{equation}
\label{c.0.1}
M_{k_1,k_2,k_3}=\int \; \dd {\bf v}\; V_x^{k_1}V_y^{k_2}V_z^{k_3} f_s({\bf V})
\end{equation}
According to the symmetry of the USF distribution $f_s$, the only nonvanishing moments correspond to
even values of $k_1+k_2$ and $k_3$. In this case, after some algebra, one gets
\beq
\label{c.0.2}
M_{k_1,k_2,k_3}=
n\left(\frac{2T}{m}\right)^{k/2}M_{k_1,k_2,k_3}^*,
\eeq
where the reduced moments $M_{k_1,k_2,k_3}^*$ are given by
\beqa
\label{c.0.3}
& & M_{k_1,k_2,k_3}^*=
\pi^{-3/2}
\sum_{\stackrel{q=0}{q+k_1=\text{even}}}^{k_1}
\frac{k_1!}{(k_1-q)!}
\Gamma\left(\frac{k_1-q+1}{2}\right)\nonumber\\
& &
\times \Gamma\left(\frac{k_2+q+1}{2}\right)
\Gamma\left(\frac{k_3+1}{2}\right)
(-\widetilde{a})^q\left(1+k \widetilde{\xi}\right)^{-(1+q)}.
\nonumber\\
\eeqa
It is easy to see that the second-degree velocity moments of the BGK model coincide with those obtained from the Boltzmann equation by using Grad's method, Eqs.\ \eqref{2.27}--\eqref{2.29}.

\section{Generalized transport coefficients}
\label{appD}

The results derived from the BGK-like kinetic model \eqref{4.1} considered to determine the generalized transport coefficients $\eta_{ijk\ell}$, $\kappa_{ij}$ and $\mu_{ij}$ are provided in this appendix. The equations defining the generalized transport coefficients in the BGK model can be obtained from Eqs.\ \eqref{3.28}--\eqref{3.30} with the replacements \eqref{4.2} and \eqref{4.3}:
\beqa
\label{c1}
& & -a c_y\frac{\partial X_{n,i}}{\partial c_x}-\left(\gamma+\frac{\zeta^{(0)}}{2}\right) \frac{\partial}{\partial{\bf c}}\cdot {\bf c}X_{n,i}+\nu \beta X_{n,i}
\nonumber\\
& & =Y_{n,i}-\frac{T}{n}\left[\frac{2a}{d p}(1-n\partial_n)
P_{xy}^{(0)}-\zeta^{(0)}\right]X_{T,i},
\eeqa
\beqa
\label{c2}
& &-a c_y\frac{\partial X_{T,i}}{\partial c_x}-\left(\gamma+\frac{\zeta^{(0)}}{2}\right) \frac{\partial}{\partial{\bf c}}\cdot {\bf c}X_{T,i}+\nu \beta X_{T,i}
\nonumber\\
& &
-\left[\frac{2a}{d n}(\partial_T P_{xy}^{(0)})+2\gamma+\frac{3}{2}\zeta^{(0)}\right]X_{T,i}=Y_{T,i},
\eeqa
\beqa
\label{c3}
& &-a c_y\frac{\partial X_{u,j\ell}}{\partial c_x}-\left(\gamma+\frac{\zeta^{(0)}}{2}\right) \frac{\partial}{\partial{\bf c}}\cdot {\bf c}X_{u,j\ell}+\nu \beta X_{u,j\ell}
\nonumber\\
& &
-a\delta_{jy}X_{u,x\ell}-\gamma X_{u,j\ell}-\frac{1}{2}\zeta_{u,j\ell}\left[\frac{\partial}{\partial{\bf c}}\cdot
({\bf c}f^{(0)})\right.\nonumber\\
& & \left.+2T\partial_Tf^{(0)}\right]=Y_{u,j\ell}.
\eeqa

In order to get the transport coefficients $\kappa_{ij}$, $\mu_{ij}$, and $\eta_{ijk\ell}$,
it is convenient to introduce the general velocity moments
\begin{equation}
\label{c4}
A_{k_1,k_2,k_3}^{(i)}=\int\, \dd {\bf c}\, c_x^{k_1}c_y^{k_2}c_z^{k_3}X_{n,i},
\end{equation}
\begin{equation}
\label{c5}
B_{k_1,k_2,k_3}^{(i)}=\int\, \dd {\bf c}\, c_x^{k_1}c_y^{k_2}c_z^{k_3}X_{T,i},
\end{equation}
\begin{equation}
\label{c6}
C_{k_1,k_2,k_3}^{(ij)}=\int\, \dd {\bf c}\, c_x^{k_1}c_y^{k_2}c_z^{k_3}X_{u,ij}.
\end{equation}
These moments provide the explicit forms of the generalized transport coefficients of the {\em perturbed} USF problem. To determine them, Eqs.\ (\ref{c1})--(\ref{c3}) are multiplied by $c_x^{k_1}c_y^{k_2}c_z^{k_3}$
and integrated over velocity. After some algebra, one achieves
\beqa
\label{c7}
& & ak_1 A_{k_1-1,k_2+1,k_3}^{(i)}+\left(\nu \beta+k \xi\right)
A_{k_1,k_2,k_3}^{(i)}+\omega_n B_{k_1,k_2,k_3}^{(i)}\nonumber\\
& & =\mathcal{A}_{k_1,k_2,k_3}^{(i)},
\eeqa
\beq
\label{c8}
ak_1B_{k_1-1,k_2+1,k_3}^{(i)}+\left(\nu \beta+k \xi+\omega_T\right)
B_{k_1,k_2,k_3}^{(i)}=\mathcal{B}_{k_1,k_2,k_3}^{(i)},
\eeq
\beqa
\label{c9}
& & ak_1C_{k_1-1,k_2+1,k_3}^{(j\ell)}+\left(\nu \beta-\gamma+k\xi \right)
C_{k_1,k_2,k_3}^{(j\ell)}+\frac{1}{2}\zeta_{u,j\ell}\nonumber\\
& & \times
\left(k-2T\partial_T\right)
M_{k_1,k_2,k_3}^{(0)}-a\delta_{jy}C_{k_1,k_2,k_3}^{(x\ell)}=
\mathcal{C}_{k_1,k_2,k_3}^{(j\ell)},
\nonumber\\
\eeqa
where $M_{k_1,k_2,k_3}^{(0)}$ are the moments of the zeroth-order distribution function $f^{(0)}$,
\beq
\label{xi}
\xi=\gamma+\frac{1}{2}\zeta^{(0)},
\eeq
\beq
\label{c10}
\omega_n= \frac{T}{n}\left[\frac{2a}{d}\left(\gamma^* P_{xy,\gamma}^*+a^* P_{xy,a}^*\right)-\zeta^{(0)}\right],
\eeq
\beq
\label{c11}
\omega_T=\frac{a}{d}\left(\gamma^* P_{xy,\gamma}^*+a^* P_{xy,a}^*\right)-\frac{1}{2}\zeta^{(0)},
\eeq
and we have introduced the quantities
\beq
\label{c11.1}
\mathcal{A}_{k_1,k_2,k_3}^{(i)}\equiv \int\, \dd {\bf c}\, c_x^{k_1}c_y^{k_2}c_z^{k_3}Y_{n,i},
\eeq
\beq
\label{c11.2}
\mathcal{B}_{k_1,k_2,k_3}^{(i)}\equiv \int\, \dd {\bf c}\, c_x^{k_1}c_y^{k_2}c_z^{k_3}Y_{T,i},
\eeq
\beq
\label{c11.3}
\mathcal{C}_{k_1,k_2,k_3}^{(j\ell)}\equiv \int\, \dd {\bf c}\, c_x^{k_1}c_y^{k_2}c_z^{k_3}Y_{u,j\ell}.
\eeq

\begin{widetext}
The integrals \eqref{c11.1}--\eqref{c11.3} can be computed with the result
\begin{eqnarray}
\label{c12}
{\cal A}_{k_1,k_2,k_3}^{(\ell)}
&=&-\frac{\partial}{\partial n}M_{k_1+\delta_{\ell x},k_2+\delta_{\ell y},k_3+\delta_{\ell z}}
+\frac{1}{\rho}
\frac{\partial P_{\ell j}^{(0)}}{\partial n}\left(\delta_{jx}k_1M_{k_1-1,k_2,k_3}+\delta_{jy}k_2M_{k_1,k_2-1,k_3}
+\delta_{jz}k_3M_{k_1,k_2,k_3-1}\right)
\nonumber\\
&=&-\left(\frac{2T}{m}\right)^{\frac{k+1}{2}}\left[
(1-\gamma^*\partial_{\gamma^*}-a^*\partial_{a^*})M_{k_1+\delta_{\ell x},k_2+\delta_{\ell
y},k_3+\delta_{\ell z}}^*\right.\nonumber\\
& &
\left.
-\frac{1}{2}(P_{\ell j}^{*}-\gamma^*P_{\ell j,\gamma}^{*}-a^*P_{\ell j,a}^{*})\left(\delta_{jx}k_1M_{k_1-1,k_2,k_3}^*+\delta_{jy}k_2M_{k_1,k_2-1,k_3}^*
+
\delta_{jz}k_3M_{k_1,k_2,k_3-1}^*\right)\right],
\end{eqnarray}
\begin{eqnarray}
\label{c13}
{\cal B}_{k_1,k_2,k_3}^{(\ell)}
&=& -\frac{\partial}{\partial T}M_{k_1+\delta_{\ell x},k_2+\delta_{\ell y},k_3+\delta_{\ell z}}+\frac{1}{\rho}
\frac{\partial P_{\ell j}^{(0)}}{\partial T}\left(\delta_{jx}k_1M_{k_1-1,k_2,k_3}+\delta_{jy}k_2M_{k_1,k_2-1,k_3}
+\delta_{jz}k_3M_{k_1,k_2,k_3-1}\right)
\nonumber\\
&=&-n\left(\frac{2T}{m}\right)^{\frac{k+1}{2}}\left[\frac{1}{2T}
(k+1-\gamma^*\partial_{\gamma^*}-a^*\partial_{a^*})M_{k_1+\delta_{\ell x},k_2+\delta_{\ell
y},k_3+\delta_{\ell z}}^*\right.\nonumber\\
& &-\frac{1}{2T}\left(P_{\ell j}^{*}-\frac{1}{2}\gamma^*P_{\ell j,\gamma}^{*}-\frac{1}{2}a^*P_{\ell j,a}^{*}\right)
\left(\delta_{jx}k_1M_{k_1-1,k_2,k_3}^*+\delta_{jy}k_2M_{k_1,k_2-1,k_3}^*\right.\nonumber\\
& & \left.\left.+
\delta_{jz}k_3M_{k_1,k_2,k_3-1}^*\right)\right],
\end{eqnarray}
\begin{eqnarray}
\label{c14}
{\cal C}_{k_1,k_2,k_3}^{(j\ell)}
&=&-\delta_{j\ell}\left(1-n\frac{\partial}{\partial n}\right)M_{k_1,k_2,k_3}+\frac{2}{d n}
\left(P_{j\ell}^{(0)}-a\eta_{xyj\ell}\right)
\frac{\partial}{\partial T}M_{k_1,k_2,k_3}\nonumber\\
& & -M_{k_1,k_2,k_3}\left(\delta_{jx}\delta_{\ell x}k_1+\delta_{jy}
\delta_{\ell y}k_2+\delta_{jz}\delta_{\ell z}k_3\right)\nonumber\\
& &
-k_1\delta_{jx}\left(\delta_{\ell y}M_{k_1-1,k_2+1,k_3}+\delta_{\ell z}M_{k_1-1,k_2,k_3+1}\right)\nonumber\\
& &
-k_2\delta_{jy}\left(\delta_{\ell x}M_{k_1+1,k_2-1,k_3}+\delta_{\ell z}M_{k_1,k_2-1,k_3+1}\right)
\nonumber\\
& &
-k_3\delta_{jz}
\left(\delta_{\ell x}M_{k_1+1,k_2,k_3-1}+\delta_{\ell y}M_{k_1,k_2+1,k_3-1}\right)\nonumber\\
&=&-n\left(\frac{2T}{m}\right)^{k/2}\left[
\delta_{j\ell} \left(\gamma^*M_{k_1,k_2,k_3,\gamma}^*+a^*M_{k_1,k_2,k_3,a}^*\right)-\frac{1}{d n T}
\left(P_{j\ell}^{(0)}-a\eta_{xyj\ell}\right)\right.
\nonumber\\
& & \times (k M_{k_1,k_2,k_3}^*
-\gamma^*M_{k_1,k_2,k_3,\gamma}^*-a^*M_{k_1,k_2,k_3,a}^*)+
M_{k_1,k_2,k_3}^* (\delta_{jx}\delta_{\ell x}k_1+
\delta_{jy}\delta_{\ell y}k_2+\delta_{jz}\delta_{\ell z}k_3)
\nonumber\\
& &
+k_1\delta_{jx}(\delta_{\ell y}M_{k_1-1,k_2+1,k_3}^*+
 \delta_{\ell z}M_{k_1-1,k_2,k_3+1}^*)+k_2\delta_{jy}(\delta_{\ell x}M_{k_1+1,k_2-1,k_3}^*+\delta_{\ell z}\nonumber\\
& &\left.\times
M_{k_1,k_2-1,k_3+1}^*)+k_3\delta_{jz}(\delta_{\ell x}
M_{k_1+1,k_2,k_3-1}^*+
\delta_{\ell y}M_{k_1,k_2+1,k_3-1}^*)\right].
\nonumber\\
\end{eqnarray}
\end{widetext}
Here, $M_{k_1,k_2,k_3}^*$ are the reduced moments of the
distribution $f^{(0)}$ defined by Eq.\ (\ref{c.0.3}). They depend on $n$ and $T$ through their dependence on $\gamma^*$ and $a^*$. In the steady state, $M_{k_1,k_2,k_3}^*$ is given by Eq.\ (\ref{c.0.3}) while the
derivatives $M_{k_1,k_2,k_3,\gamma}^*$ and $M_{k_1,k_2,k_3,a}^*$ can be obtained by
following the procedure described in appendix \ref{appA}.

The solution to Eqs.\ \eqref{c7}--\eqref{c9} can be written as
\beqa
\label{c15}
A_{k_1,k_2,k_3}^{(i)}&=&(\nu \beta)^{-1}\sum_{q=0}^{k_1} \frac{k_1!}{(k_1-q)!}
(-\widetilde{a})^q\left(1+k\widetilde{\xi}\right)^{-(1+q)}\nonumber\\
& & \times\left(
{\cal A}_{k_1-q,k_2+q,k_3}^{(i)}-\omega_n B_{k_1-q,k_2+q,k_3}^{(i)}\right),
\nonumber\\
\eeqa
\beqa
\label{c16}
B_{k_1,k_2,k_3}^{(i)}&=&(\nu \beta)^{-1}\sum_{q=0}^{k_1} \frac{k_1!}{(k_1-q)!}
(-\widetilde{a})^q\nonumber\\
& & \times \left(1+\widetilde{\omega}_T
+k\widetilde{\xi}\right)^{-(1+q)} {\cal B}_{k_1-q,k_2+q,k_3}^{(i)},
\nonumber\\
\eeqa
\begin{eqnarray}
\label{c17}
C_{k_1,k_2,k_3}^{(j\ell)}&=&(\nu \beta)^{-1}\sum_{q=0}^{k_1} \frac{k_1!}{(k_1-q)!}
(-\widetilde{a})^q \nonumber\\
& & \times
\left(1-\widetilde{\gamma}
+k\widetilde{\xi}\right)^{-(1+q)}\left[ {\cal
C}_{k_1-q,k_2+q,k_3}^{(j\ell)}\right.\nonumber\\
& & +a\delta_{jy}C_{k_1-q,k_2+q,k_3}^{(x\ell)}
-\frac{1}{2}n\left(\frac{2T}{m}\right)^{k/2}\zeta_{u,j\ell}\nonumber\\
& & \times \left.\left(\gamma^*M_{k_1-q,k_2+q,k_3,\gamma}^*
+a^*M_{k_1-q,k_2+q,k_3,a}^*\right)
\right],
\nonumber\\
\end{eqnarray}
where $\widetilde{\omega}_T\equiv \omega_T/(\nu \beta)$. In Eqs.\ \eqref{c15}--\eqref{c17}, we recall that in the steady state the parameter $\widetilde{\xi}=\chi$ is given by Eq.\ \eqref{2.30}, and $\gamma^*= \beta \chi-\frac{1}{2}\zeta_0^*$. The expressions of the generalized transport coefficients $\kappa_{ij}$, $\mu_{ij}$ and $\eta_{ijk\ell}$ can be obtained from Eqs.\ \eqref{c15}--\eqref{c17}.


\begin{thebibliography} {55}
\bibitem{note}As noted in Ref.\ \cite{G06}, the original kinetic model proposed in Ref.\ \cite{BDS99} employed the exact local homogeneous cooling state instead of the (local) equilibrium distribution function $f_\text{M}$. On the other hand, given that the latter distribution is an excellent approximation to the former one in the region of thermal velocitites, it is expected that the use of $f_\text{M}$ provides reliable results for transport around USF.
\bibitem{code}A Mathematica code has been included as a suplementary material to evaluate the generalized transport coefficients plotted in Figs.\ \ref{fig5.1}--\ref{fig9} as functions of the (reduced) sher rate $a^*$ and the coefficient of restitution $\al$.
\end{thebibliography}
\end{document}